\documentclass{emulateapj}
\usepackage{makebox}
\usepackage{graphicx,scalerel}
\usepackage{mathrsfs} 
\usepackage{xcolor}
\newcommand{\vecv}{\mbox{\boldmath $v$} {}}
\newcommand{\vecr}{\mbox{\boldmath $r$} {}}

\newcommand{\RSNe}{\mbox{$R_{\rm \scriptscriptstyle SNe}$}{}}
\newcommand{\ESNe}{\mbox{$E_{\rm \scriptscriptstyle SNe}$}{}}
\newcommand{\MSNe}{\mbox{$M_{\rm \scriptscriptstyle SNe}$}{}}
\newcommand{\MBH}{\mbox{$M_{\rm \scriptscriptstyle BH}$}{}}
\newcommand\wh[1]{\hstretch{2}{\hat{\hstretch{.5}{#1}}}}
\newcommand{\smalls}{\scriptscriptstyle}

\shorttitle{Supernova explosions in accretion disks}
\shortauthors{Moranchel-Basurto et al.}

\begin{document}

\title{Supernova explosions in accretion disks in active galactic
nuclei: Three-dimensional models}

\author{A. Moranchel-Basurto\altaffilmark{1},
F. J. S\'anchez-Salcedo\altaffilmark{2},
Ra\'ul O. Chametla\altaffilmark{3},
and P. F. Vel\'azquez\altaffilmark{1}}

\altaffiltext{1} {Instituto de Ciencias Nucleares, Universidad Nacional Aut\'onoma de M\'exico, CP 04510. Mexico City, Mexico}
\altaffiltext{2}{Instituto de Astrono\'ia,
Universidad Nacional Aut\'onoma de M\'exico, Ciudad Universitaria, Apt. Postal 70 264, C.P. 04510, Mexico City, Mexico}
\altaffiltext{3}{Instituto de Ciencias F\'isicas, Universidad Nacional Aut\'onoma de M\'exico, Av. Universidad s/n, 62210 Cuernavaca, Mor., Mexico}

\begin{abstract}

Supernova (SN) explosions can potentially affect the structure and evolution of
cirumnuclear disks in active galactic nuclei (AGN). Some previous studies have
suggested that a relatively low rate of SN explosions can provide an effective value 
of alpha viscosity between $0.1$ and $1$ in AGN accretion disks within $1$ pc scale.
In order to test this possibility, we provide some analytic scalings of the evolution 
of a SN remnant embedded in a differentially rotating smooth disk.  
We calibrate our estimates using three-dimensional hydrodynamical simulations 
where the gas is modeled as adiabatic with index $\gamma$. Our simulations 
are suited to include the fact that a fraction of the momentum injected 
by the SN escapes from the disk into the 
corona. Based on these results, we calculate the 
contribution of SN explosions
to the effective alpha viscosity, denoted by $\alpha_{\smalls \rm SNe}$, in a model AGN accretion disk, where accretion is driven by the local viscosity $\alpha$. 
We find that for AGN galaxies with a central black hole of $\sim 10^{8}M_{\odot}$
and a disk with viscosity $\alpha=0.1$, the contribution of SN explosions may be 
as large as $\alpha_{\smalls \rm SNe}\simeq 0.02$, provided that $\gamma\gtrsim 1.1$.  
On the other hand, in the momentum conservation limit, which is valid when
the push by the internal pressure of the SN remnant is negligible, 
we find $\alpha_{\smalls \rm SNe} \lesssim 6\times 10^{-4}$. 

\end{abstract}


\section{Introduction}

It is well established that the luminosity of the active galactic nuclei (AGN) is the 
result of gas being accreted by the central supermassive black hole (SMBH). 
In order to reach the central SMBH, the ISM gas must 
be transported from galactic scales down to the last stable orbit.
At galactic scales,  gravitational torques produced by bars and arms can bring the gas into 
the central sub-kpc region \citep[e.g.,][]{shl89,shl90,hop11}.
At intermediate radii (scales $1$-$100$ pc), the circumnuclear disk or torus, 
is gravitationally unstable. As a result, part of the gas turns into stars.
It has been suggested that energy feedback from supernovae (SN)
inside the circumnuclear disks may support a geometrically thick, turbulent 
circumnuclear disk \citep[e.g.,][]{wad02,kaw08}.
\citet{hob11} suggest that feedback from stellar 
winds and SN within a kpc-scale disk can make the gas highly turbulent 
at ``intermediate scales'' and this turbulence can promote accretion
\citep[see also][]{hop12}.

At the pc and sub-pc scales from the nucleus, models predict the formation
of a thin accretion disk surrounding the SMBH. Interestingly, reverberation
mapping indicates that broad line regions (BLRs) in AGNs have a characteristic radial distance
from $0.01$ to $1$ pc \citep[e.g.,][]{kas07,ben09,du19},
implying that BLRs overlap with the outer parts of the accretion disks. 
Understanding the interplay between the underlying accretion
disk, the BLR and the inner parts of the dusty torus is crucial in order to 
have a complete picture of the physical processes.

Models suggest that the accretion disk is gravitationally unstable
between $\sim 0.01$ pc and a few pc from the SMBH,
and thus it is likely to fragment into clouds leading to a star-forming disk 
\citep[e.g.,][]{pac78,shl87,col99,col01,goo03}.
Vigorous star formation could raise material from the surface 
of the underlying disk, feeding the BLR with metal-rich gas \citep{wan12,cze16}
If so, the BLR can trace the metallicity of the accretion disk \citep{wan11}.

Emission-line flux ratios have been used to estimate 
the BLR metallicity. It has been found that, in many quasars, 
the BLR metallicity is very high, with typical values of $4-5$ times solar 
\citep[e.g.,][]{ham93,bal03,die03,war03,nag06,kur07,jia07,jua09}.
The almost constant BLR line ratios over the redshift range suggests that such
high metallicities were reached by a rapid and intense chemical enrichment at the
cores of the host galaxies. However, the fact that the metallicity of host galaxies
is generally lower than it is in BLRs, indicates that the BLRs have been enriched locally 
within the star-forming disk, rather than by the stellar population of the host galaxy
\citep{wan11}.

A major long-standing puzzle is to explain how gas can power the central SMBH 
if a significant fraction of the gas in the radial inflow is consumed in forming stars.
The problem of the consumption of gas can be alleviated if the angular momentum
transport in the disk is efficient, i.e. if the effective $\alpha$ viscosity is $\gtrsim 0.1$
\citep{shl89,goo03,che09,wan10,wan11}. 
Some authors have suggested that  
repeated SN explosions in the star-forming disk, produce
an effective viscosity $\alpha\sim 0.1$ \citep{roz95, col99, col08,che09,wan10,wan11}. 
However, there is still no consensus
on the rate of SN explosions in the accretion disk required to provide 
$\alpha \sim 0.1$. The most detailed study on the amount of angular momentum
redistributed by a single SN explosion in the accretion disk is
given in \citet{roz95}. Based on two-dimensional (2D) simulations,
\citet{roz95} showed that the blast wave driven by a SN explosion deflects the trajectory
of disk gas elements, resulting in an outward angular momentum flux,
because the mixing between elements that have
acquire angular momentum with those that have lost angular momentum is small.
They argued that a SN rate as low as $10^{-4}$ yr$^{-1}$ in
a disk rotating around a black hole of $(10^{8}-10^{9})M_{\odot}$ provides a radial flux of 
angular momentum corresponding to a viscosity parameter $\alpha\sim 0.1$.

\citet{col99,col08} built a steady-state accretion disk, including
star formation feedback. They argued that the angular momentum redistributed
by one SN is lower than the value derived by \citet{roz95}.
Therefore, they require a larger number of
SN explosions to have the same rate of angular momentum transport.
The difference between the prescriptions of \citet{roz95} and \citet{col99,col08} is related to the uncertainties on the amount of momentum 
that escapes from the disk carried by the outflowing gas
when the SN remnant (SNR) breaks out of the disk.

In the present paper, we reconsider the angular momentum transport in
the AGN accretion disks provided by SN explosions. In Section \ref{sec:basis}, we describe 
the physical model of the disk. Section \ref{sec:SN_basic} 
gives a general insight into the evolution of SNR in
accretion disks around SMBH. In particular, we provide estimates about
conditions for the breakout of the disk and give scaling laws
for the radial width of the cavity opened by a single SN explosion and
the redistribution of angular momentum. In Section \ref{sec:simulations}, 
we present the results of three-dimensional (3D) simulations, which take into 
account that some fraction of the mass, energy and momentum can be carried 
outside the disk by the vertical outflow induced by the explosion. In Section \ref{sec:repeated_SN}, we apply our results to estimate the effective $\alpha$ 
viscosity induced by SN explosions.
Finally, a summary of the main conclusions is given in Section \ref{sec:conclusions}.

\section{Model and disk parameters}
\label{sec:basis}

We consider an accretion disk in the gravitational field of a central SMBH with
mass $M_{\rm \smalls BH}$ and Schwarzschild radius $R_{\rm Sch}$.  
In the outer regions of accretion disks, i.e. beyond $10^{3}R_{\rm Sch}$,
the dust sublimates and the opacity drops.
Consequently, these outer regions are expected to be gravitationally
unstable and fragmentation of the disk into clouds seems unavoidable 
\citep[e.g.,][]{goo03,raf09,jia11}.
It is usually assumed that feedback from newly formed stars can maintain the 
outer parts of accretion disks marginally stable, so that the Toomre $Q$-parameter 
remains close to $1$, inhibiting further star formation
(e.g., Collin \& Zhan 1999; Gammie 2001; Goodman 2003; 
Sirko \& Goodman 2003, hereafter SG; 
Thompson et al. 2005, hereafter TQM; Rafikov 2009; Begelman \& Silk 2017). 
Radiation pressure from massive stars and from the accretion of gas
onto stellar black holes in the disk, momentum injection 
from SN explosions, thermal pressure as well as magnetic fields can
provide the support necessary to maintain $Q\simeq 1$.

There is a wide variety of plausible theoretical models to describe the
structure of quasi-stationary self-gravitating accretion disks. They differ
on the assumed speed of the radial inflow, the adopted disk opacities and on
the importance of the magnetic support. 
For instance, in the SG model, accretion is driven by local turbulent 
viscosity, which is parametrized by the Shakura-Sunyaev viscosity parameter $\alpha$.
TQM presume that a global torque could
be able to drive a larger inflow speed than local viscosity.
As a result, at distances from the SMBH between $10^{3}R_{\rm Sch}$ 
and $10^{5}R_{\rm Sch}$,
the disk surface density and thickness are smaller in the TQM model
than they are in the SG model (a comparison between the outcomes of these
models is given in Figure 1 in \citet{bel16}, assuming plausible parameters).
Other models suggest that magnetic fields play an important role on the
gas dynamics. For example, in magnetically elevated disks, the inflow, which is
carried by low-density gas at large heights,
is driven mainly by stress due to the large-scale component
of the magnetic field \citep{mis20}.

The angular momentum transport due to local turbulence by 
magnetorotational instabilities (MRI) and gravitational instabilities (GI) can
be accomodated in the $\alpha$ model. The effect of SN explosions on the disk
can also contribute to enhance the effective viscosity, which
can be also described by a viscosity parameter $\alpha_{\smalls \rm SNe}$.
The $\alpha$ parameter will be the sum of all the above contributions:
\begin{equation}
\alpha=\alpha_{\smalls \rm MRI}+\alpha_{\smalls \rm GI}+\alpha_{\smalls \rm SNe}+
\alpha_{\smalls \rm others},
\label{eq:alpha_contributions}
\end{equation}
where $\alpha_{\smalls \rm others}$ represent the viscosity due to other potential
sources, e.g., star-disk collisions \citep{par03}. The viscosity coefficients in
the right hand side of Eq. (\ref{eq:alpha_contributions}) are not independent.
In the present work, we estimate the contribution of SN explosions to the angular
momentum transport, i.e. $\alpha_{\smalls \rm SNe}$, in a disk model
where the local viscosity $\alpha$ drives the inflow.
Self-consistent models require $\alpha_{\smalls \rm SNe}\lesssim \alpha$.
In those models where $\alpha_{\smalls \rm SNe}\sim \alpha$,  SN explosions by
themselves can account for the rate of angular momentum transfer.

In the SG model, accretion is driven by local viscosity. In this model,
the inflow mass rate $\dot{M}_{\rm acc}$ is constant with radius and
assumed to be $l_{\rm Edd}L_{\rm Edd}/(\eta c^{2})$, where $l_{\rm Edd}$ is the 
Eddington ratio and $\eta$ the radiative efficiency.  In terms of the
dimensioneless parameters $\alpha_{\smalls 0.1}=\alpha/0.1$
and $\xi= l_{\rm Edd}/(\alpha_{\smalls 0.1}\eta_{\smalls 0.1})$, it can be written as
\begin{equation}
\dot{M}_{\rm acc}=2.2 \alpha_{\smalls 0.1} \xi M_{8} \; M_{\odot} {\rm yr}^{-1},
\label{eq:dotMacc}
\end{equation}
where $M_{8}=\MBH/(10^{8}M_{\odot})$. 
Typical ranges for the values of these parameters are: $l_{\rm Edd}=0.1-0.5$,
$\eta_{\smalls 0.1}\simeq 1$ and $\alpha_{\smalls 0.1}=0.1-3$.

Between a radius $\sim 10^{3}R_{\rm Sch}$ and the 
outer radius $10^{5}R_{\rm Sch}$, Equation (\ref{eq:dotMacc}) 
together with the condition $Q=Q_{\rm m}$ (where $Q_{\rm m}\simeq 1$) determine 
the surface density $\Sigma (R)$, the scale height $H(R)$, the midplane total pressure
$p_{0}(R)$ and the effective sound speed $c_{s}(R)$ (including
thermal, radiation and turbulent pressure). More specifically, the Toomre parameter
in a Keplerian disk can be expressed as:
\begin{equation}
Q=\frac{H \Omega^{2}}{\pi G \Sigma} = \frac{\Omega^{2}}{\sqrt{2\pi^{3}} G \rho_{0}},
\label{eq:Qparameter}
\end{equation}
where $\Omega(R)$ is the angular velocity of the disk. Note that in the second equality 
of Equation (\ref{eq:Qparameter}),
we have used that the midplane density in a vertically isothermal 
disk is $\rho_{0}=\Sigma/(\sqrt{2\pi}H)$. The condition $Q\simeq Q_{\rm m}$
implies $H= \pi Q_{\rm m} G \Sigma /\Omega^{2}$,
$c_{s}=\Omega H= \pi Q_{\rm m} G\Sigma/\Omega$ and 
\begin{equation}
\rho_{0}=\frac{\Omega^{2}}{\sqrt{2\pi^{3}} Q_{\rm m}G}.
\end{equation}
The midplane total pressure is
$p_{0}=\rho_{0} c_{s}^{2}=\sqrt{\pi/2} Q_{\rm m}G\Sigma^{2}$.

\begin{figure}
\includegraphics[scale=0.38]{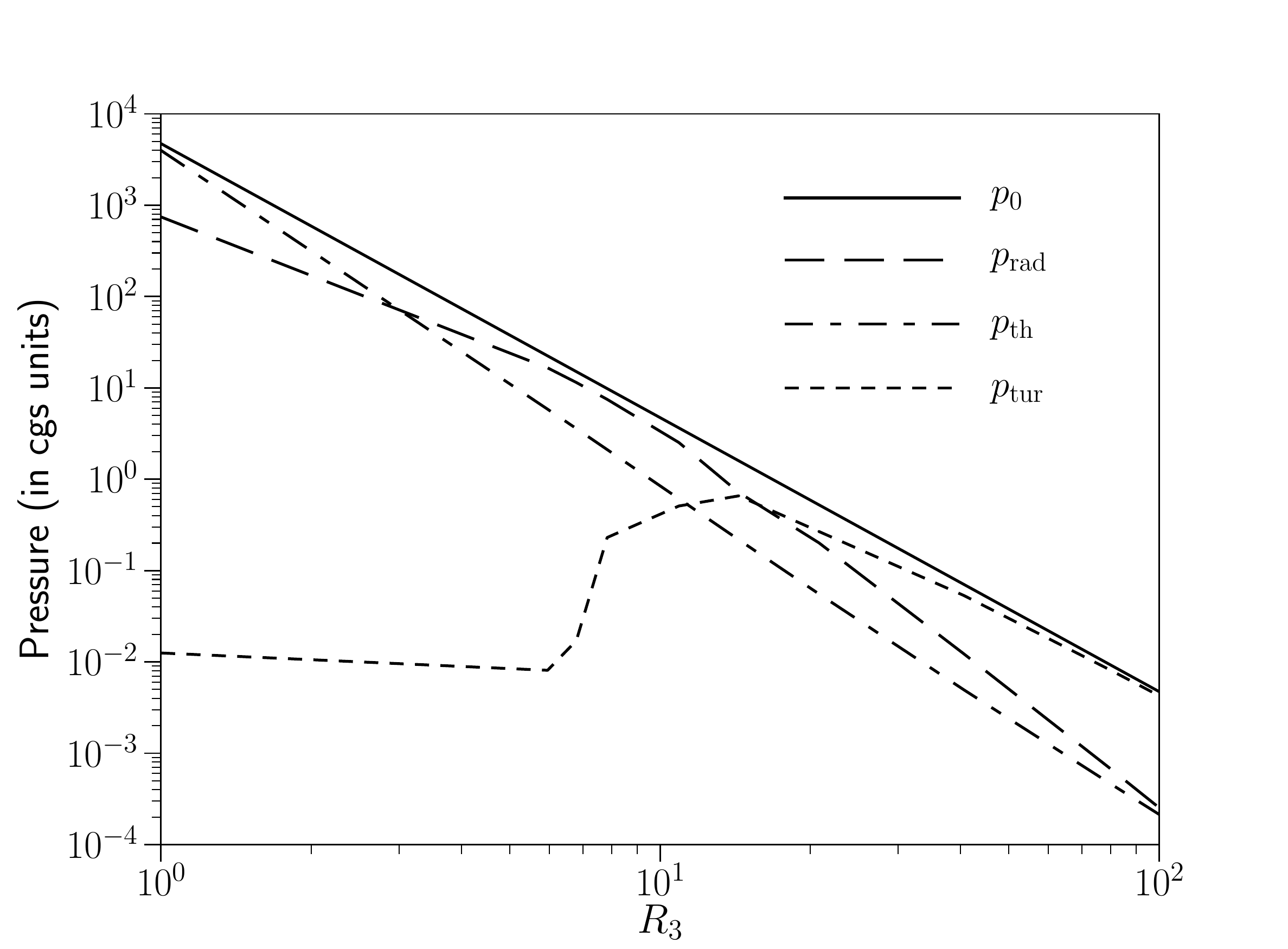}
\caption{Radial profile for the different components of pressure in a model with
$M_{8}=1$, $\alpha_{\smalls 0.1}=1$, $\xi=0.3$ and $Q_{\rm m}=1$.} 
\label{fig:pressure_profile}
\vskip 0.3cm
\end{figure}

By imposing the inflow mass rate given in Equation (\ref{eq:dotMacc}),
the SG model predicts
\begin{equation}
\Sigma(R)= 1.6\times 10^{9} \xi^{1/3} Q_{\rm m}^{-2/3}
M_{8}^{-2/3} R_{3}^{-3/2} M_{\odot} {\rm pc}^{-2},
\label{eq:initial_Sigma}
\end{equation}
\begin{equation}
H(R)=4.6\times 10^{-5} \xi^{1/3} Q_{\rm m}^{1/3}M_{8}^{4/3} R_{3}^{3/2} {\rm pc},
\label{eq:initial_H}
\end{equation}
and 
\begin{equation}
c_{s}=  32 \xi^{1/3} Q_{\rm m}^{1/3}M_{8}^{1/3} \, {\rm km} \; {\rm s}^{-1},
\label{eq:vrms}
\end{equation}
where $R_{3}=R/(10^{3}R_{\rm Sch})$.
The aspect ratio of the disk $h(R)\equiv H/R$ is 
\begin{equation}
h(R)= 4.6\times 10^{-3} \xi^{1/3} Q_{\rm m}^{1/3}M_{8}^{1/3} R_{3}^{1/2}.
\end{equation}
For $Q_{\rm m}=1$, $M_{8}=1$ and $\xi=0.3$, the disk is very thin, 
with an aspect ratio 
$h=3.1\times 10^{-3}$ at $R_{3}=1$. If we take $\xi=5$ as in SG, $h=8\times 10^{-3}$
at $R_{3}=1$, in agreement with the value plotted in Figure 2 in SG. 
These values for the aspect ratio are intermediate between those found in the TQM model
($h\sim 10^{-3}$, at $R_{3}\sim 1$) and those in magnetically elevated disks
($h\sim 0.05$, Mishra et al. 2020).

The scale height given in Equation (\ref{eq:initial_H}) 
includes all the components that provide support to the disk:
the thermal pressure ($p_{\rm th}$), the radiation pressure ($p_{\rm rad}$)
and the turbulent pressure\footnote{In the
ongoing discussion, we are not considering magnetic fields.} ($p_{\rm tur}$).
The latter is thought to be mostly produced by
random motions in the gas excited by SN explosions and stellar winds. 
Following TQM (see Appendix \ref{sec:TQM_appendix} for details), 
we have computed the different contributions
($p_{\rm th}$, $p_{\rm rad}$ and $p_{\rm tur}$) to the total pressure.
The approach of TQM rests on the assumption that star formation is the main 
feedback mechanism.
In that scenario, the radiation-to-turbulent pressure ratio is 
$\simeq \tau/2$, where $\tau$ is the midplane optical depth. 
Figure \ref{fig:pressure_profile} shows the pressure distribution in a disk with  
$Q_{\rm m}=1$, $\alpha_{\smalls 0.1}=1$, $\xi=0.3$ and $M_{8}=1$.
We see that the inner region ($R_{3}\lesssim  2.5$) is dominated by thermal pressure, 
an intermediate region dominated by radiation 
pressure  ($2.5 \lesssim R_{3}\lesssim 15$),
and a region dominated by turbulent pressure at 
$15\lesssim R_{3}\lesssim 100 $.

As noted by TQM, a potential problem of disk models driven by a local viscosity 
is that the star formation rate is so large that it consumes almost all the gas,
and little gas is left to fuel the central BH (see also \S \ref{sec:SNe_area}). 
However, there are many uncertainties regarding the feedback. In a scenario where
irradiation of the disk can also occur from the
accretion of gas onto intermediate-mass black holes 
embedded in the disk \citep{dit20}, the ratio between radiation to
turbulent pressure could be much larger than $\tau/2$. The SG disk 
is considered a plausible model and has been adopted to 
study the migration of stellar-mass black holes within AGN disks \citep{mck12,bel16,sec19,yan20}.

Througout this paper we will make discussions using the SG model 
(Eqs. \ref{eq:initial_H}-\ref{eq:vrms}). But since we provide analytical formulae for 
calculating the impact of SN explosions on the disk,
the interested reader can apply them to other disk models.

\section{SN in AGN accretion disks: Evolution of a single explosion}
\label{sec:SN_basic}
For simplicity, we consider the evolution of a SN that explodes in the midplane of the accretion disk, at a radial distance $\RSNe$ from the SMBH, in the range
$10^{3}R_{\rm Sch}\leq \RSNe\leq 10^{5}R_{\rm Sch}$. The explosion has
an energy $\ESNe$ and a mass ejecta $M_{\rm \smalls SNe}$.
The momentum associated to this energy and mass is 
$P_{\rm \smalls SNe}=(2M_{\rm \smalls SNe}\ESNe)^{1/2}$. 
The SN explosion will drive a shock in the disk and will carve a cavity of 
low-density gas, surrounded by a shell of swept material. 
As a first approach to the evolution of a SN explosion in the disk,
we do not include the physics of radiation pressure, radiative transport
and self-gravity. Although our hydrodynamical approach is clearly simplistic,
we believe that it captures much of the physics of interest.

Given that the problem is multidimensional, it is useful to have analytical
estimates of the key quantities.  In this Section, we make predictions
regarding the radial extent of the cavity,
and the redistribution of angular momentum of the disk by a single SN explosion
{in an initially unperturbed disk with surface density and half thickness as 
given in Eqs. (\ref{eq:initial_Sigma})-(\ref{eq:initial_H}).}
As it will become clear in \S \ref{sec:repeated_SN}, we need these quantities to
characterize the contribution of SN explosions to the 
effective viscosity coefficient.

\subsection{First stage: free expansion phase}
\label{sec:free_expansion}
An initial free expansion phase will take place until a mass comparable to the
SN ejecta has been swept up. 
Let $\Delta$ denote the radius of a cylinder perpendicular to the disk at the position
$\RSNe$ that contains a mass $M_{\rm \smalls SNe}$. 
By its definition 
$\Delta^{2} \equiv M_{\rm \smalls SNe}/(\pi \tilde{\Sigma})$,
where the tilde over a quantity indicates evaluation at $R_{\rm \smalls SNe}$,
i.e. $\tilde{\Sigma}\equiv \Sigma(R_{\rm \smalls SNe})$.
In terms of $\tilde{H}$, it is
\begin{equation}
\frac{\Delta}{\tilde{H}}= \xi^{-1/2} M_{8}^{-1} \wh{M}_{10}^{1/2}  
\tilde{R}_{3}^{-3/4},
\end{equation}
where $\tilde{R}_{3}=R_{\rm \smalls SNe}/(10^{3}R_{\rm Sch})$ and
$\wh{M}_{10}=M_{\rm \smalls SNe}/(10M_{\odot})$.
Thus, for 
\begin{equation}
\tilde{R}_{3}>\xi^{-2/3} M_{8}^{-4/3} \wh{M}_{10}^{2/3},
\end{equation}
we have $\Delta \lesssim \tilde{H}$. 
This condition implies that when the SNR has a radius $\sim \tilde{H}$,
the SNR has swept an amount of matter comparable to or larger than the
SN ejecta. In particular, for $M_{8}\geq1$, $\wh{M}_{10}=1$
and $\xi=0.3$, this condition is met if the
SN explosion occurs at a radius $\tilde{R}_{3}$ larger than $2.2$.

In order to make the presentation clearer, we will consider in the next 
Sections \ref{sec:breakout} and \ref{sec:width} only situations in which 
$\Delta \lesssim \tilde{H}$. The case $\Delta > \tilde{H}$ is presented 
in the Appendix \ref{sec:limit1}.

\subsection{Conditions for breakout of the disk}
\label{sec:breakout}

A fraction of the momentum and energy released by the SN vent into the corona 
if the SNR can breakout of the disk. In this subsection we consider the conditions
in which breakout of the disk occurs. For simplicity, we will implicitly assume that 
$\Delta$ is significantly smaller than $\tilde{H}$.

\citet{kom60} studied the propagation of the shock wave in a 
plane-parallel stratified medium. It was shown that if the explosion
occurs at the midplane, the SNR is prolate in shape
because the expansion velocity of the shell 
will be higher in the vertical direction than in the radial direction.
Let denote $Z_{\rm sh}(t)$ the vertical distance of the topmost point of the shell
(i.e., the semimajor axis in the $z$-direction of the prolate SNR)
and $R_{\rm sh}(t)$ the radius of the SNR in the $z=0$ plane.
$\dot{Z}_{\rm sh}$ decreases with $z$ up to a certain height $z_{b}$, and then it is 
reaccelerated \citep[e.g.,][]{kom60,mac89,fer00}. 
The SN explosion will break out of the disk if $v_{b}>\tilde{c}_{s}$ where $v_{b}$ 
is the velocity of the shock $\dot{Z}_{\rm sh}$ at $z_{b}$ \citep[e.g.,][]{fer00}.

In order to calculate $v_{b}$, we consider two limiting scenarios. In scenario A,
we assume that the SNR reaches the reaccelerating height $z_{b}$ with negligible
radiative cooling.
Thus, we may use the Kompaneets approximation to evaluate
the $z$-component of the velocity of the topmost point of the shell as
$\dot{Z}_{\rm sh}(z) \simeq [\ESNe/(\tilde{\rho} z^{3})]^{1/2}$, where
$\tilde{\rho}(z)=\tilde{\rho}_{0}\exp(-z^{2}/[2\tilde{H}^{2}])$
\citep[e.g.,][]{fer00,ola09}.
Scenario B assumes that cooling is already important at the end
of the free expansion phase so that the pressure interior to the shell is negligible.
In such a case, momentum-conservation 
implies $\dot{Z}_{\rm sh}(z) \simeq 3P_{\rm \smalls SNe}/(4\pi \tilde{\rho} z^{3})$,
again if $Z_{\rm sh}<z_{b}$. 
Other possibilities that may be considered more realistic, as a pressure modified
snowplow phase, lie between these two limiting scenarios.

Assuming that the disk has a vertical Gaussian
profile, we find $z_{b}=\sqrt{3}\tilde{H}$ in both scenarios.
In scenario A and for our disk model (Eqs. \ref{eq:initial_Sigma} and 
\ref{eq:initial_H}), we get
\begin{equation}
v_{b}=5.5\times 10^{3} \xi^{-1/2} M_{8}^{-1}  E_{51}^{1/2}\tilde{R}_{3}^{-3/4}
{\rm km}\, {\rm s}^{-1},
\end{equation}
where  
$E_{51}= E_{\rm \smalls SNe}/(10^{51} {\rm erg})$. The breakout 
condition $v_{b}>\tilde{c}_{s}$ implies that SNR can punch a hole in the disk
if the explosion occurs within a radius $\tilde{R}_{b}^{(A)}$ given by
\begin{equation}
\tilde{R}_{3}\leq \tilde{R}_{b}^{\smalls (A)} \equiv
1.0\times 10^{3} \xi^{-10/9} M_{8}^{-14/9} E_{51}^{2/3}.
\label{eq:breakout_A}
\end{equation}
For the reference values $\xi=0.3$ and $E_{51}=1$, we find that 
$\tilde{R}_{b}^{\smalls (A)}\gtrsim 20$, if $M_{8}\leq 30$. In particular, for $M_{8}=1$,
we obtain that $\tilde{R}_{b}^{\smalls (A)}= 3.8\times 10^{3}$, which is much larger than the outer edge of the accretion disk. 

In scenario B, we get
\begin{equation}
v_{b}\simeq 4.5\times 10^{3} \xi^{-1} M_{8}^{-2}  \wh{M}_{10}^{1/2}
E_{51}^{1/2}\tilde{R}_{3}^{-3/2}
{\rm km}\, {\rm s}^{-1}.
\end{equation}
The breakout condition is fulfilled if 
\begin{equation}
\tilde{R}_{3}\leq \tilde{R}_{b}^{\smalls (B)} \equiv
30 \,\xi^{-8/9} M_{8}^{-14/9} \wh{M}_{10}^{1/3}E_{51}^{1/3}.
\label{eq:breakout_B}
\end{equation}
For $\xi=0.3$ and $M_{8}=1$, a SN explosion with $E_{51}=\wh{M}_{10}=1$
is capable to break out of the disk in the range
of interest $1\leq \tilde{R}_{3}\leq 100$.
On the contrary, if $M_{8}\geq 18$, SN explosions with $E_{51}=\wh{M}_{10}=1$
will never breakout of the disk at $\tilde{R}_{3}>1$.

\subsection{Radial width of the SNR}
\label{sec:width}
In the $z=0$ plane, the otherwise circular SNR will be deformed 
by the Coriolis forces in a first stage,
when the expansion velocity of the SNR $\dot{R}_{\rm sh}$ is 
still much larger than the shear velocity.
In a second stage, when $\dot{R}_{\rm sh}$ becomes comparable to the shear
velocity, the SNR will become elongated along the azimuthal direction, resembling
an ellipse in shape, due mainly to differential rotation induced by the ram pressure 
with the ambient medium \citep{ola82,ten87,pal90,sil92,roz95}.
In a final stage, shear dominates and the radial width of the cavity $W$ may decrease
over time \citep[e.g.,][]{ten87}.
In this section, we evaluate
the maximum radial width of the SNR, denoted by $W_{\rm max}$,
in scenario A.
An analogue derivation but for scenario B can be found in Appendix \ref{sec:scenarioB}.

Let denote $W_{\rm max}^{\smalls \rm trans}$ the width of the SNR 
when the shock velocity drops to a value similar to the effective sound speed of the 
external medium (i.e. the shock velocity becomes transonic), and 
$W_{\rm max}^{\smalls \rm shear}$ the radial width of the SNR
when the shock velocity is comparable to the
shear velocity $\simeq (3/4)\tilde{\Omega} R_{\rm sh}$, where $\Omega$ is
the angular velocity.
The maximum width of the SNR will be given by $W_{\rm max}
={\rm min}\left[W_{\rm max}^{\smalls \rm trans}, 
W_{\rm max}^{\smalls \rm shear}\right]$.

The expansion velocity $\dot{R}_{\rm sh}$ of the SNR depends on the thermodynamics 
of the gas and on the fraction of momentum that is transferred into the disk. 
In order to compute $\dot{R}_{\rm sh}$, we will consider the adiabatic phase and the momentum-driven snowplow phase. Therefore, we implicitly assume that
a mass of gas larger than the SN ejecta has been swept up.
As a consequence, our results are valid only if 
$\tilde{\rho}_{0} W_{\rm max}^{3}\gtrsim 8M_{\smalls \rm SNe}$, i.e.
$W_{\rm max}>W_{\rm lim}\equiv 2(M_{\smalls \rm SNe}/\tilde{\rho}_{0})^{1/3}$.

Consider first the case where the SN explosion occurs at a distance 
$\tilde{R}_{3}<\tilde{R}_{b}^{\smalls (A)}$ from the SMBH. If so, the SN can 
make a hole in the disk. At short times after the explosion, before breakout of the disk,
when $R_{\rm sh}^{2}+Z_{\rm sh}^{2}\ll \tilde{H}^{2}$, the shell is spherical because
the SNR evolves as if the medium were homogeneous, so that $\dot{R}_{\rm sh}=
\dot{Z}_{\rm sh}\simeq (2/5)(\ESNe/[\tilde{\rho}_{0}R_{\rm sh}^{3}])^{1/2}$.
The interior pressure of the SNR continue pushing
the shell in both the vertical and radial directions while the SNR is embedded 
within the disk, i.e. while $R_{\rm sh}\sim Z_{\rm sh}\lesssim \tilde{H}$.
At some point, the flow accelerates in the vertical direction leading
to a rapid depressurization of the cavity. This phase of depressurization starts when
$R_{\rm sh}\simeq \tilde{H}$ and $Z_{\rm sh}\simeq \sqrt{3}\tilde{H}$ (see \S
\ref{sec:breakout}). We recall here that $\tilde{H}$ is the scaleheight of the disk prior breakout, which includes the vertical support provided by the turbulent pressure. 
Anticipating to what we observe in the simulations, the SNR forms
an overdense ring-like structure in the disk, with expansion velocity $\dot{R}_{\rm sh}$.
If we assume that after depressurization of the cavity,
the ring-like SNR enters into a snowplow phase,
momentun conservation may be expressed as
\begin{equation}
R_{\rm sh}^{2} \dot{R}_{\rm sh}\simeq \chi_{\smalls A} \tilde{H}^{2} \dot{R}_{\rm sh}(\tilde{H}),
\label{eq:Rsh_scenarioA}
\end{equation}
where $\chi_{\smalls A}$ is a dimensionless correction factor to account for the push by
the interior pressure during the pressure-driven phase.
Equation (\ref{eq:Rsh_scenarioA}) encapsulates the fact 
that the interior pressure
of the cavity can push the shell for a longer time if the disk (prior to breakout)
is thick, imparting more momentum to the shell than in more thin disks 
\citep[e.g.,][]{ten87}.
From the Kompaneets approximation, we have
$\dot{R}_{\rm sh}(\tilde{H})\simeq 0.4(\ESNe/[\tilde{\rho}_{0}\tilde{H}^{3}])^{1/2}$.
Using this value in Equation (\ref{eq:Rsh_scenarioA}), we can derive $\dot{R}_{\rm sh}$.
The transonic condition implies
\begin{equation}
W_{\rm max}^{\smalls A, \rm trans}\simeq 
1.2\left(\frac{\chi_{\smalls A}^{2} \ESNe \tilde{H}}
{\tilde{\rho}_{0} \tilde{c}_{s}^{2}}\right)^{1/4}.
\label{eq:Wmax_A_trans_inner}
\end{equation}
In our disk model
\begin{equation}
W_{\rm max}^{\smalls A, \rm trans}\simeq 7.5\times 10^{-4} \chi_{\smalls A}^{1/2}
\xi^{-1/12 } Q_{\rm m}^{1/6}E_{51}^{1/4} M_{8}^{2/3} \tilde{R}_{3}^{9/8} \, {\rm pc}.
\label{eq:Wmax_A_trans_model1}
\end{equation}
The condition $\dot{R}_{\rm sh}\simeq (3/4)\tilde{\Omega} R_{\rm sh}$ leads to
\begin{equation}
W_{\rm max}^{\smalls A,\rm shear}\simeq 
1.6\left(\frac{\chi_{\smalls A}^{2}\ESNe \tilde{H}}{\tilde{\rho}_{0} \tilde{\Omega}^{2}}\right)^{1/6}.
\label{eq:Wmax_A_shear_inner}
\end{equation}
The dependence of $W_{\rm max}^{\smalls A, \rm shear}$ on 
$\ESNe$, $\tilde{\rho}_{0}$ and $\tilde{H}$ is weak. This agrees with the fitting formula
found in \citet{pal90} for the SNR minor axis in the $z=0$ plane.
Using a model in $1.5$ dimensions that describes the propagation of the shell
from a strong explosion in a rotating disk,
they found that the semiminor axis is proportional to 
$\ESNe^{0.22} \tilde{\rho}_{0}^{-0.22} \tilde{H}^{0.1}$ (their Equation 4).
In terms of our independent variables and for our disk model, Equation (\ref{eq:Wmax_A_shear_inner}) becomes
\begin{equation}
W_{\rm max}^{\smalls A,\rm shear}\simeq 4\times 10^{-4} 
\chi_{\smalls A}^{1/3}\xi^{1/18} Q_{\rm m}^{2/9}M_{8}^{8/9} E_{51}^{1/6} 
\tilde{R}_{3}^{5/4} {\rm pc}.
\label{eq:Wmax_A_shear_inner_model1}
\end{equation}

\begin{figure}
\includegraphics[scale=0.42]{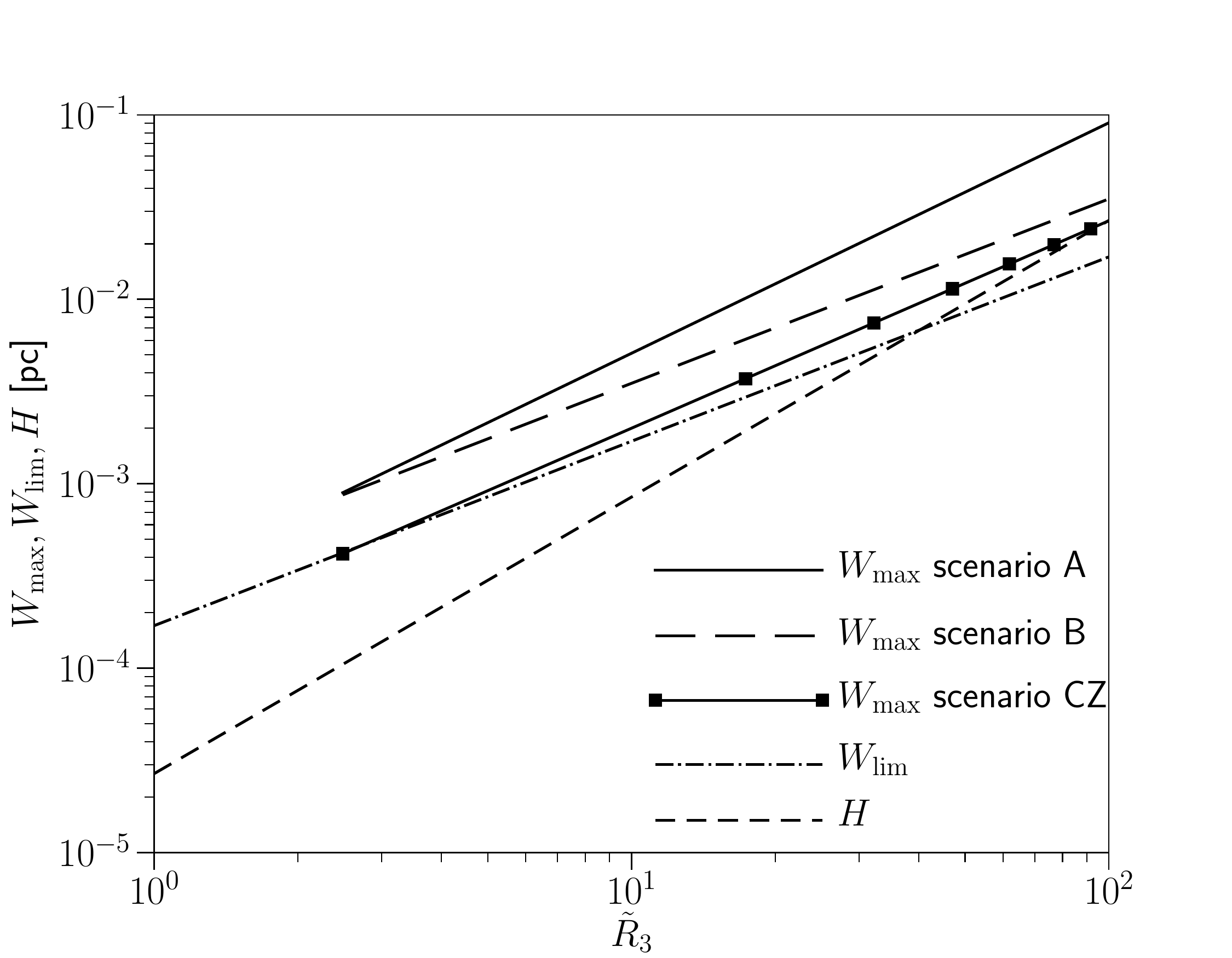}
\caption{$W_{\rm max}$ as a function of radius for a model with $M_{8}=1$ 
in scenario A (solid line), in scenario B (long dashed line) and using Eq. (\ref{eq:Wmax_CZ}) (solid line with small squares). We take $\xi=0.3$, $Q_{\rm m}=1$, $E_{51}=1$
and $\chi_{\smalls A}=\chi_{\smalls B}=1$. The scale height of
the disk $H$, and $W_{\rm lim}\equiv 2(M_{\smalls \rm SNe}/\tilde{\rho}_{0})^{1/3}$ are also shown. For this range 
of parameters, the SNR breaks out of the disk.} 
\label{fig:Wmax_M8_1}
\vskip 0.3cm
\end{figure}

If the explosion occurs at $\tilde{R}_{3}>\tilde{R}_{b}^{\smalls (A)}$,
the disk is not perforated. Thus, the shell is confined to the disk.
When the explosion occurs at the midplane, Kompaneets's approximation implies
that $\dot{R}_{\rm sh}\simeq 0.4 (\ESNe/[\tilde{\rho}_{0}R_{\rm sh}^{3}])^{1/2}$
if $Z_{\rm sh}\leq z_{b}=\sqrt{3}\tilde{H}$
\citep[see Fig. 2 in][]{ola09}.
The transonic condition
$\dot{R}_{\rm sh}\simeq \tilde{c}_{s}$, leads to
\begin{equation}
W_{\rm max}^{\smalls A,\rm trans}\simeq 
\left(\frac{\ESNe}{\tilde{\rho}_{0} \tilde{c}_{s}^{2}}\right)^{1/3}.
\end{equation}
For our accretion disk model (Eqs. \ref{eq:initial_Sigma}-\ref{eq:vrms}), 
we can recast $W_{\rm max}^{\smalls A, \rm trans}$ in 
terms of $M_{8}$ and $\tilde{R}_{3}$ as
\begin{equation}
W_{\rm max}^{\smalls A, \rm trans}\simeq 1.5\times 10^{-3} \xi^{-2/9} Q_{\rm m}^{1/9}
E_{51}^{1/3} M_{8}^{4/9} \tilde{R}_{3} \, {\rm pc}.
\end{equation}
On other hand, if the radial expansion
of the SNR is limited by shear,
the condition $\dot{R}_{\rm sh}\simeq (3/4)\tilde{\Omega} R_{\rm sh}$ implies
\begin{equation}
W_{\rm max}^{\smalls A,\rm shear}\simeq 0.8 
\left(\frac{\ESNe}{\tilde{\rho}_{0} \tilde{\Omega}^{2}}\right)^{1/5}.
\end{equation}
For our disk model,
\begin{equation}
W_{\rm max}^{\smalls A,\rm shear}\simeq 6\times 
10^{-4} Q_{\rm m}^{1/5} M_{8}^{4/5} E_{51}^{1/5} 
\tilde{R}_{3}^{6/5} {\rm pc}.
\end{equation}

The corresponding values of $W_{\rm max}$ in scenario B are given in the Appendix
\ref{sec:scenarioB} (see Eqs. \ref{eq:Wmax_trans_B}-\ref{eq:Wmax_shear_B}).
Figure \ref{fig:Wmax_M8_1} compares $W_{\rm max}$ in scenario A and 
scenario B, assuming $Q_{\rm m}=1$, $M_{8}=1$ and $E_{51}=1$.
The values of the width $W_{\rm max}$ are larger than $\tilde{H}$ because the 
SNR breakouts of the disk, for the range of values of $\tilde{R}_{3}$ under
consideration. Since the SNR in scenario B evolves as a pure
momentum-driven snowplow, the values of $W_{\rm max}^{\smalls B,\rm trans}$
and  $W_{\rm max}^{\smalls B,\rm shear}$
are smaller than the corresponding values in scenario A.
We should note that we have assumed $\chi_{\smalls A}=\chi_{\smalls B}=1$
for convenience. 

\begin{figure}
\includegraphics[scale=0.42]{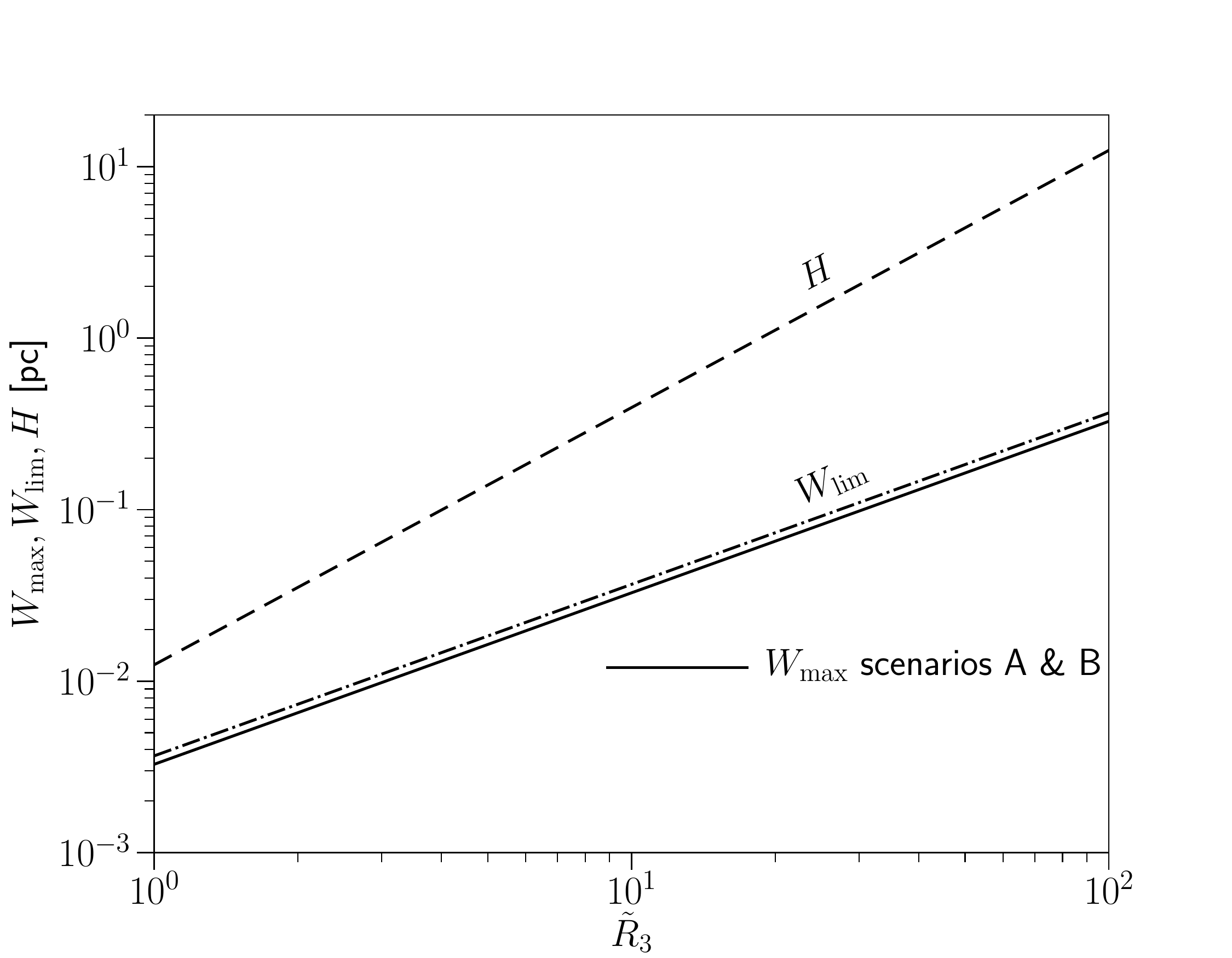}
\caption{$W_{\rm max}$ as a function of radius for a model with $M_{8}=100$ 
in scenario A and B (solid line). We take $\xi=0.3$, $Q_{\rm m}=1$ and $E_{51}=1$.
The scale height of the disk $H$ and $W_{\rm lim}$ are also shown. For this range 
of parameters, the SNR is confined to the disk.} 
\label{fig:Wmax_M8_100}
\vskip 0.2cm
\end{figure}

Figure \ref{fig:Wmax_M8_100} shows the same quantities
as Figure \ref{fig:Wmax_M8_1} but
for $M_{8}=100$. SNR formed within the range $10^{3}R_{\rm Sch}$ and 
$10^{5}R_{\rm Sch}$ and $E_{51}=1$, have not enough energy to break out of the
disk.  In addition, we see that the curve of $W_{\rm max}$ is very close to
the curve for $W_{\rm lim}$. This indicates that the values of $W_{\rm max}$ should
be taken with caution. $W_{\rm max}$ in scenarios A and 
B are similar. In fact, the curves for scenarios A and B overlap.

There has been other attempts to estimate $W_{\rm max}$ in the literature. \citet{col99}
suggest that for SN explosions powerful enough as to produce breakout of the disk,
the momentum transferred to the disk is 
$P_{\rm \smalls SNe} W_{\rm max}/(2\tilde{H})$.
As a result, they find 
\begin{equation}
W_{\rm max}^{\smalls \rm CZ, shear}\approx 1.1\left(\frac{P_{\rm \smalls SNe}}{\pi \tilde{\rho}_{0} \tilde{\Omega}}\right)^{1/4}.
\label{eq:Wmax_CZ}
\end{equation}
For the parameters of Figure \ref{fig:Wmax_M8_1}, $W_{\rm max}^{\smalls \rm CZ, shear}$
is roughly a factor of $2-3$ smaller than $W_{\rm max}^{\smalls \rm B,shear}$.
This implies that we would need $\chi_{\smalls B}\simeq 0.1$ to have a match 
between both estimates.
3D simulations can shed light on the appropriate
value for the fraction of momentum that is absorbed by the disk.

\subsection{Angular momentum redistribution in the disk by SN explosions}
\label{sec:ang_L_redistribution}
SN explosions may redistribute the angular momentum of the disk. As discussed
in \citet{roz95}, disk gas elements that enter the shock facing the inner
disk lose angular momentum, whereas gas elements that cross the shock facing
the outer disk gain angular momentum. The net effect is a transport of angular
momentum radially outwards. 
Denote ${\mathcal{J}}_{\rm \smalls SNe}$
the total amount of angular momentum that is transported
from the inner disk to the outer disk by just one SN explosion.
\citet{roz95} show that 
$\mathcal{J}_{\rm \smalls SNe}$ is given by
\begin{equation}
\mathcal{J}_{\rm \smalls SNe}=\mu \tilde{\Sigma} \dot{R}_{\rm sh} R_{\rm sh}^{2}
\RSNe,
\label{eq:delta_J_o}
\end{equation}
where $R_{\rm sh}$ and $\dot{R}_{\rm sh}$ should be evaluated 
once the SNR has breakout of the disk
(see Section 4.2 in \citet{roz95}).
In Equation (\ref{eq:delta_J_o}) we have introduced the fudge factor $\mu$ to be 
fixed through our numerical 
simulations in Section \ref{sec:simulations}. This factor may depend on the scenario, 
so we will refer to them as $\mu_{\smalls A}$ and $\mu_{\smalls B}$.

In scenario A, we can use Eqs. (\ref{eq:Rsh_scenarioA}) and
(\ref{eq:delta_J_o}), plus the disk scaling laws in 
Eqs. (\ref{eq:initial_Sigma}) and (\ref{eq:initial_H}) to obtain
\begin{equation}
\mathcal{J}^{\smalls (A)}_{\rm \smalls SNe}=
0.6 \beta_{\smalls A} (\ESNe \tilde{\Sigma})^{1/2}\tilde{H} \RSNe
=75 \beta_{\smalls A}\xi^{1/2} E_{51}^{1/2} M_{8}^{2}\tilde{R}_{3}^{7/4}
\label{eq:JJ_single_A}
 \end{equation}
where $\beta \equiv \mu \chi$ and
$\mathcal{J}^{\smalls (A)}_{\rm \smalls SNe}$ is in units of
$M_{\odot} {\rm pc} \,{\rm km/s}$.

In scenario B, momentum conservation implies
$\dot{R}_{\rm sh}R_{\rm sh}^{2}=\beta_{\smalls B} P_{\rm \smalls SNe}/(\pi \tilde{\Sigma})$ (see Eq. \ref{eq:scenarioB_dotR_R2}). Thereby, we find
\begin{equation}
\mathcal{J}^{\smalls (B)}_{\rm \smalls SNe}=\frac{\beta_{\smalls B}}{\pi} 
P_{\rm \smalls SNe}\RSNe = 100 \beta_{\smalls B}M_{8} \wh{M}_{10}^{1/2} E_{51}^{1/2}
\tilde{R}_{3},
\label{eq:Delta_J_roz}
\end{equation}
again in units of $M_{\odot} {\rm pc} \,{\rm km/s}$.
The next Section is devoted to simulate
the 3D evolution of a SNR in a disk and inferences of $\beta$ will be
provided.

\section{Simulations}
\label{sec:simulations}
Our 3D simulations of the evolution of a SNR in a disk in Keplerian 
rotation were performed using the code {FARGO3D\footnote{The code is publicly available at http://fargo.in2p3.fr}}
\citep{Ben2016} in a spherical coordinate system $(r,\theta,\phi)$. Magnetic fields
 and self-gravity of the disk were ignored.

We placed the site of the SN explosion at the midplane of the disk. Given the symmetry
of problem, we simulated only the upper half of the disk.
We chose a system of reference that rotates with the angular velocity at $\RSNe$, 
so that the explosion site does not change over time. We took $M_{8}=1$ and 
$\tilde{R}_{3}=20$, which corresponds to $\RSNe=0.2$ pc. In this model, the 
circular velocity of a test particle with orbital radius $\RSNe$ is $1467$ km s$^{-1}$ 
and its orbital period $P_{\rm orb}$ is $838$ yr.

The initial surface density is given in Equation (\ref{eq:initial_Sigma}).
At $\RSNe$, it is $1.2\times 10^{7}M_{\odot}$pc$^{-2}$.
The initial vertical profile of density was derived by assuming that
the temperature of the gas is independent of $\theta$, and imposing 
hydrostatic equilibrium with an aspect ratio $h=0.01375 (R/\RSNe)^{1/2}$. 
Thus the isothermal sound speed is $21.4$ km s$^{-1}$, constant along the disk.

\begin{figure}
\includegraphics[scale=0.34]{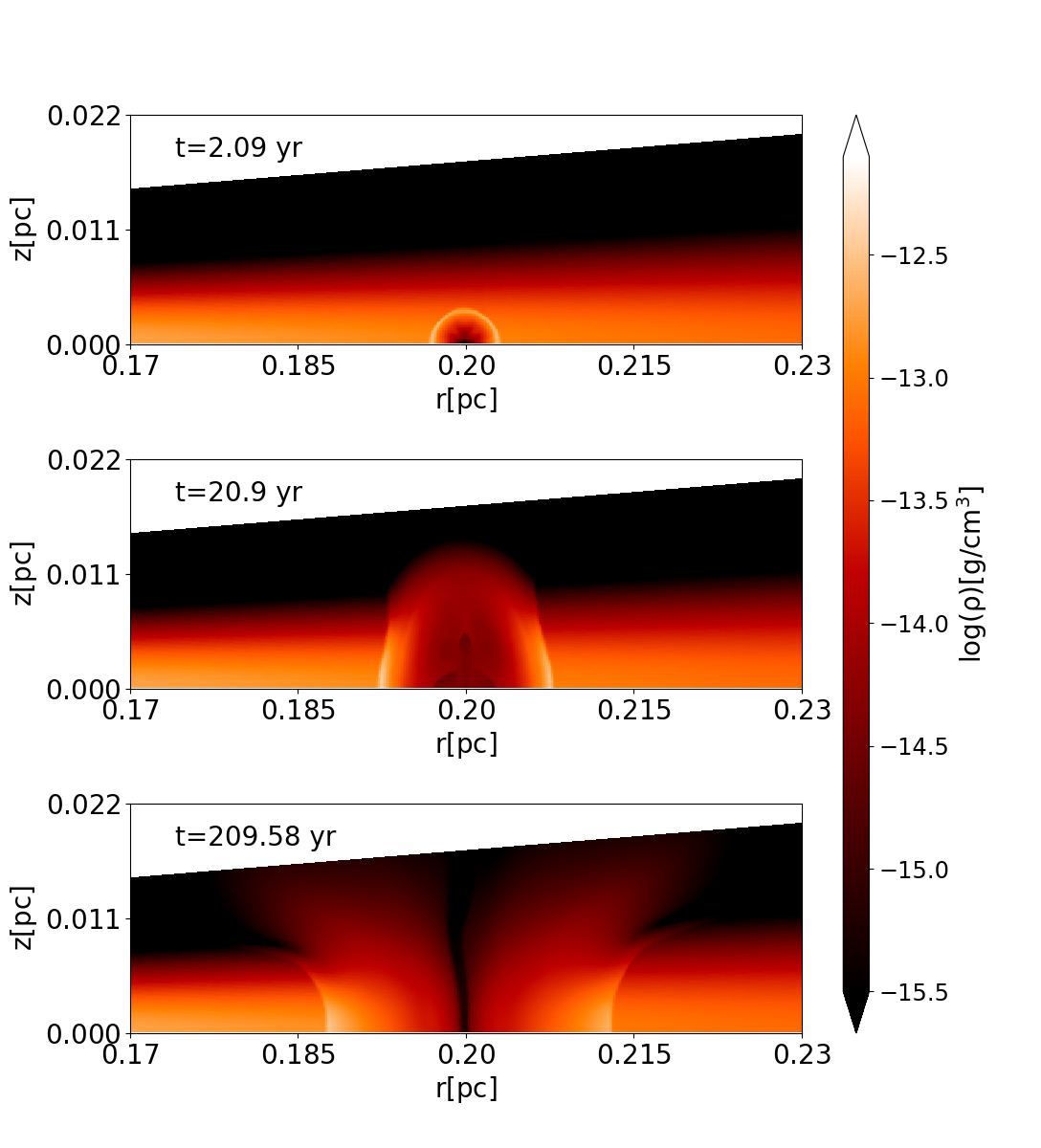}
\caption{Vertical cut of the gas density along the radial direction, passing through
the SN explosion site ($\phi=\pi/4$), in model 1.}
\label{fig:zcut_1}
\end{figure}

\begin{figure*}
\includegraphics[width=17.7 cm, height=17.75 cm]{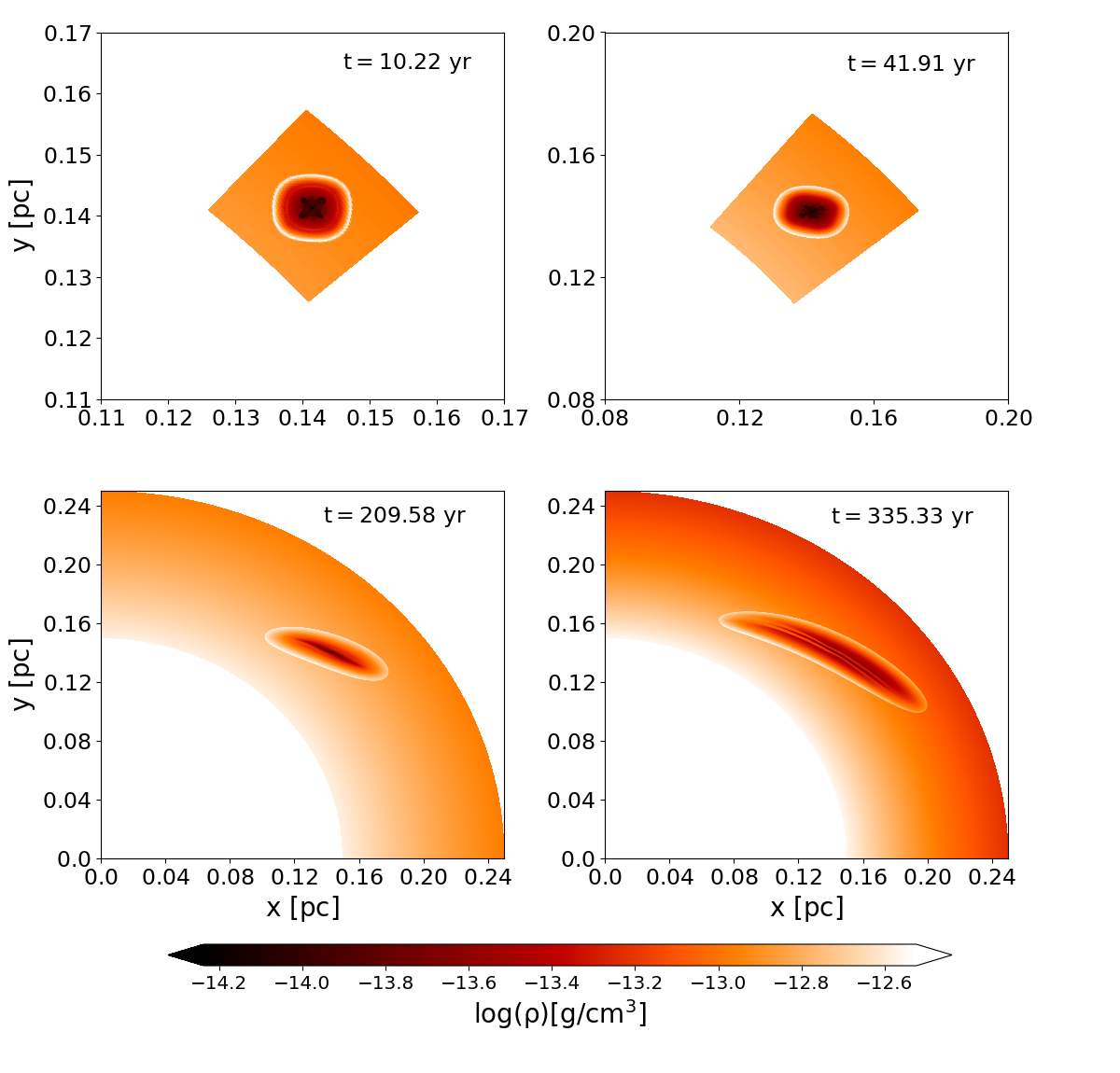}
\caption{Volume density at the midplane ($z=0$) in model 1, at four different times.} 
\label{fig:plane_dens_1}
\end{figure*}

The mass ejecta is $\MSNe=10 M_{\odot}$ and the explosion has an 
energy $\ESNe=2\times 10^{51}$ erg. As we are modeling only one half of the disk, 
we deposit $5M_{\odot}$ and $10^{51}$ erg in our domain. 
At $t=0$, this mass and energy is injected by increasing the density and the 
thermal energy of the gas into a region with a radius of $6.5\times 10^{-5}$ pc. The equation of energy is solved by
assuming the equation of ideal gas $p=(\gamma-1)e$ where $p$ is the gas pressure 
and $e$ the internal energy density.
For the adiabatic index, we take $\gamma=5/3$ (model 1) and $\gamma=1.1$ 
(model 2).

The azimuthal angle ranges from $0$ to $\pi/2$. Hence, we only
simulate an octant of the disk. The explosion center is placed at $\phi=\pi/4$.
The latitude, $\pi/2-\theta$, ranges
from $0$ (midplane of the disk) to $6.2\tilde{h}$ (in radians),
where $\tilde{h}$ is the aspect ratio of the disk at the
explosion center. In the radial direction,
the domain extends from $r_{\rm in}=0.15$ pc to $r_{\rm out}=0.25$ pc. 
In the upper tap of the disk we employ open boundary conditions. Damping 
boundary conditions for the radial component of the velocity has been used at 
$r_{\rm in}$ and at $r_{\rm out}$ \citep{dev06}.
The number of zones in each direction are $N_{r}=768$, $N_{\theta}=128$ and $N_{\phi}=2304$.

\subsection{Evolution of the SNR}
We will first focus on model 1. Model 2 will be discussed at the end of this Section.
Since model 1 is adiabatic with $\gamma=5/3$,
we will use scenario A, which ignores cooling, to make predictions.

For the parameters in model 1, the condition for breakout of the disk 
(Equation \ref{eq:breakout_A}) is satisfied.
Figure \ref{fig:zcut_1} shows vertical cuts of the volume density 
along the radial direction, passing through the explosion center, in model 1.
At $t=20.9$ yr, the disk is in the process of breakout; the blast wave 
in the $z$-direction is just leaving the disk. The shape of the SNR
in this vertical plane is prolate 
($R_{\rm sh}=2.7\tilde{H}$ and $Z_{\rm sh}=4.85\tilde{H}$, at $t=20.9$ yr)
because the $z$-direction presents less resistance.
Before the breakout of the disk, 
the shell in these cuts appears as convex arcs, but after breakout these
arcs become concave.
Some of the vented
gas escapes from our computational domain. 
Before presenting a more quantitative analysis of the mass lost through
the boundaries of our computational domain, we will look at the evolution
of the SNR along different cuts.

\begin{figure}
\includegraphics[width=9.3 cm, height=7.4 cm]{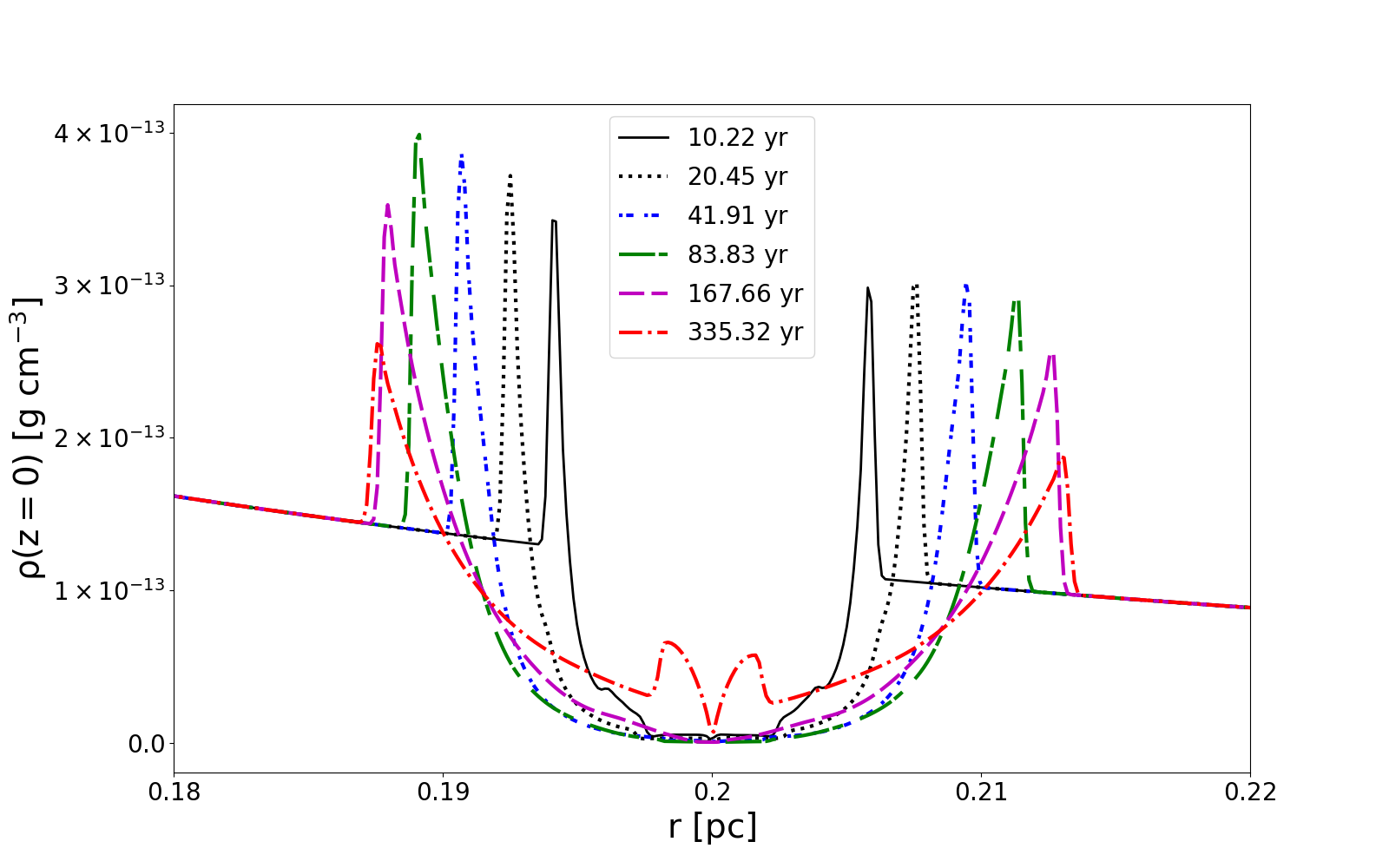}
\caption{Volume density at the midplane of the disk as a function of radius $r$,
in the direction of the explosion center ($\phi=\pi/4$), at different times, in model 1.}
\label{fig:rho_radial_midplane_model1_5first}
\end{figure}

The evolution of the volume density in the midplane of the disk
is shown in Figure \ref{fig:plane_dens_1}.  The gas disk rotates counterclockwise.
At $t\leq 20$ yr, the SNR along a cut through the $z=0$ plane
is almost circular, though a bit elongated due to the Coriolis force.
At $t=41.91$ yr, the SNR has an ellipsoidal
shape with axis ratio $\sim 0.72$. The angle between the major axis of the
ellipse and the radial direction is $\simeq 45^{\circ}$.
Due to the differential rotation of the accretion disk, the shear continues 
stretching the SNR in the azimuthal direction (see panel at $t=209.58$ yr). 
At this time, the SNR presents a banana-like shape.
Between $t=209.6$ yr ($0.25 P_{\rm orb}$) and $t=335.3$ yr ($0.4 P_{\rm orb}$),  
the major axis of the SNR continues growing, while its minor axis (width along 
the radial direction) barely changes. 

\begin{figure}
\includegraphics[width=9.3cm,height=7.3cm]{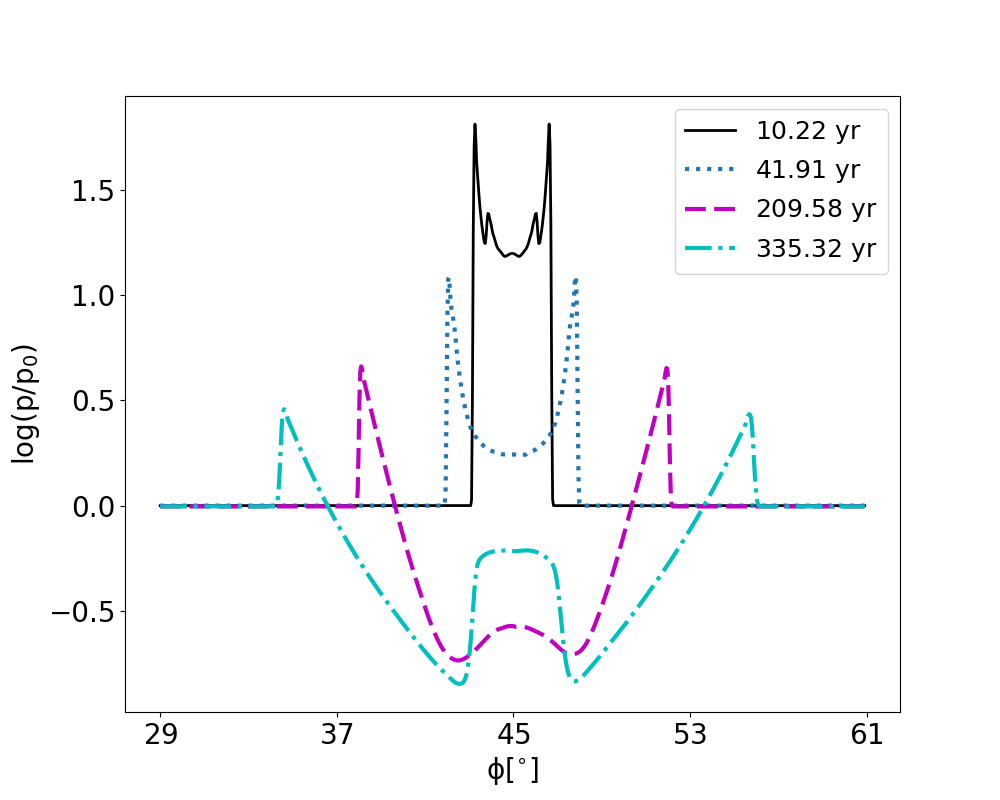}
\caption{Cross sections of the gas pressure along the azimuthal direction
at $z=0$ and $R=\RSNe$ in model 1. $p_{0}$ is the unperturbed pressure. Note that $6^{\circ}$ are $0.021$ pc.} 
\label{fig:pressure_azimuth_model1}
\end{figure}

\begin{figure}
\includegraphics[scale=0.55]{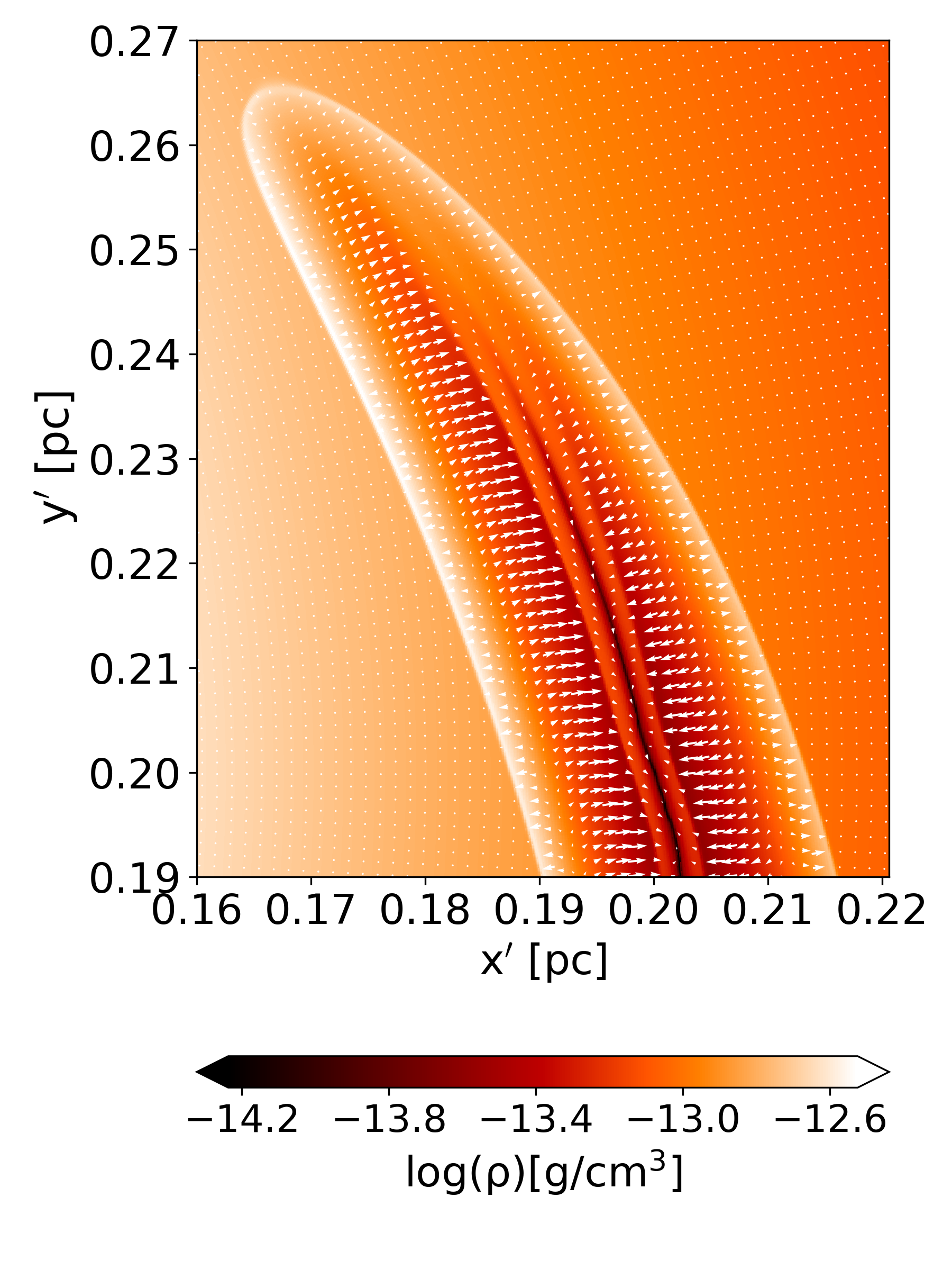}
\caption{Color map of the volume density overlayed with the perturbed velocity field 
(after substracting the Keplerian velocity field) at $z=0$ and $t=335.3$ yr in model 1.}
\label{fig:vel_vec_model1}
\end{figure}

The overall evolution of the shape of the SNR in the disk midplane is similar to that described in \citet{ten87}
in the context of formation of holes in
galactic H\,{\sc i} disks by the explosion of multiple SN in evolved OB associations. 
In such a study, they used a $1.5$-dimensional model of the SNR and assumed a flat 
rotation curve \citep[see also][]{pal90}.
The morphology of the SNR is also similar to the results in
\cite{roz95}, who made 2D simulations of a SN explosion in a Keplerian disk.

To gain a more physical insight to the evolution of the SNR, 
Figure \ref{fig:rho_radial_midplane_model1_5first} shows the volume gas density 
at the midplane of the disk along a radial cut that passes through the explosion center,
i.e. for $\phi=\pi/4$. As expected, at the very 
center of the SNR, the material has been evacuated efficiently.
The maximum radial width of the SNR is $W_{\rm max}=0.026$ pc or, in terms
of the local scale height $W_{\rm max}=9.5\tilde{H}$.
Between $10.2$ yr and $83.8$ yr, the peaks in density, which correspond to the position of the shell,
are equally spaced, implying an effective radial expansion velocity in km s$^{-1}$
\begin{equation}
\dot{W}= 5\times 10^{3} \exp \left(-\frac{W}{0.005}\right),
\end{equation}
where $W$ is in pc.
The effective velocity $\dot{W}/2$ is supersonic at $t\leq 100$ yr.
We also see that during the first $100$ yr after the explosion, the SNR is
able to keep the cavity clean of material, indicating that the SNR is able to deflect
the disk gas entering the SNR.

\begin{figure}
\includegraphics[height=7.2cm,width=9.2cm]{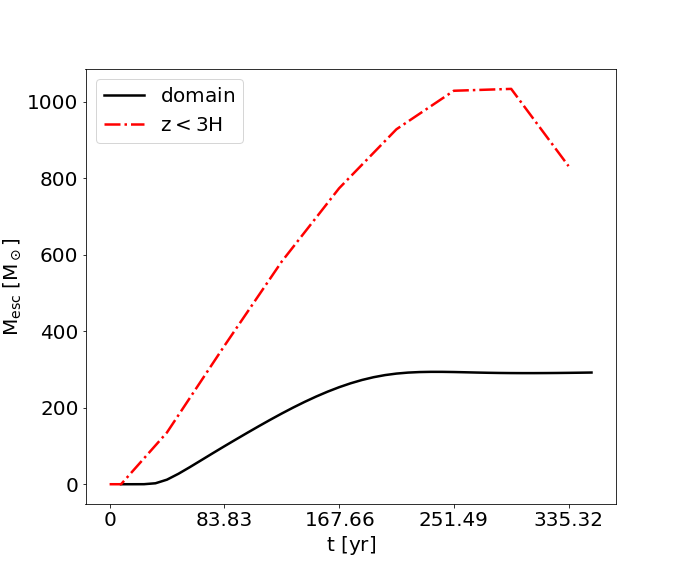}
\caption{Mass that has escaped from our computational domain (solid line) and
the mass that has exited the region $z<3H$ (dashed line), in model 1.} 
\label{fig:lost_mass_model1}
\end{figure}

After $t\simeq 100$ yr, the behaviour of the SNR changes.
$W$ increases very slowly, so that the expansion of the SNR in the radial direction has almost stalled at $t=167$ yr (see Figure \ref{fig:rho_radial_midplane_model1_5first}).
After $t=167$ yr, the density in the cavity starts to increase.
The reason is that the disk gas can penetrate into the cavity. In fact, as time goes on,
the shock waves weaken and the pitch angle decreases. As a result, the deflection
of the disk gas that enters the SNR is much more moderate.
The elements inside the cavity are accelerated inward due to the pressure gradient
that tries to refill the cavity (see Figure \ref{fig:pressure_azimuth_model1}).
Indeed, after $100$ yr, 
the central cavity is depressurized and the evolution follows a 
momentum-conservation phase.
The disk velocity field at $335.3$ yr after the explosion is shown in 
Figure \ref{fig:vel_vec_model1}. In this ``passive'' phase, the SNR only grows along
the azimuthal direction.

\begin{figure}
\includegraphics[width=8.5cm,height=8cm]{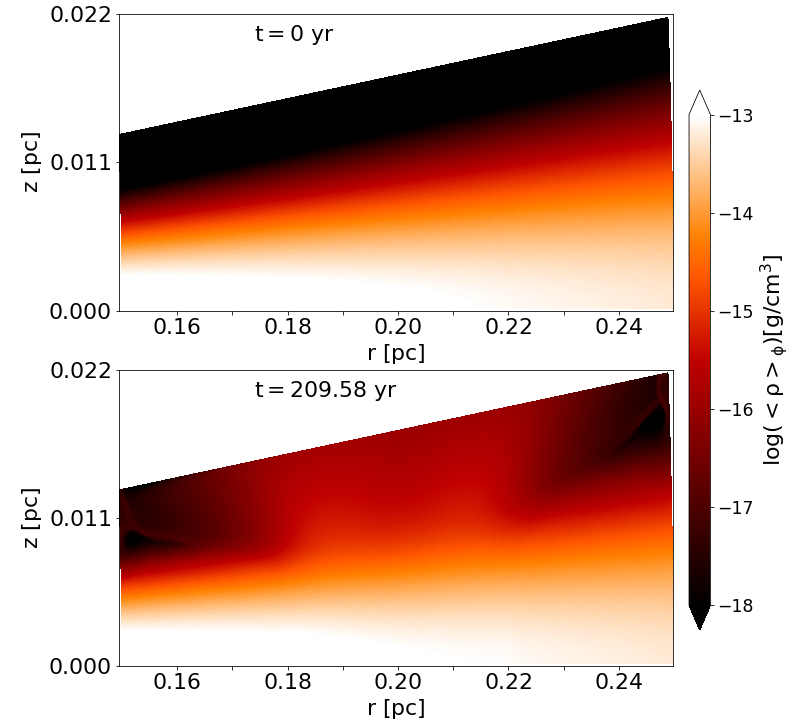}
\caption{Azimuthally-averaged volume density in model 1 at $t=0$ (top) and
at $t=209.58$ yr (bottom).}
\label{fig:density_averaged_phi}
\end{figure}

Since our simulations are 3D, we have information
about the redistribution of mass in the vertical direction, and on the 
amount of mass that is vented into the corona. 
Figure \ref{fig:lost_mass_model1} 
shows the mass that has been lost from our computational domain at a given time.
The mass loss rate
is approximately constant between $t=40$ yr and $t=170$ yr.  Mass expulsion from our
domain is halted at $200$ yr after the explosion. Approximately a mass of $300M_{\odot}$ is lost through the upper 
boundary.

\begin{figure}
\includegraphics[width=9.2cm,height=6.9cm]{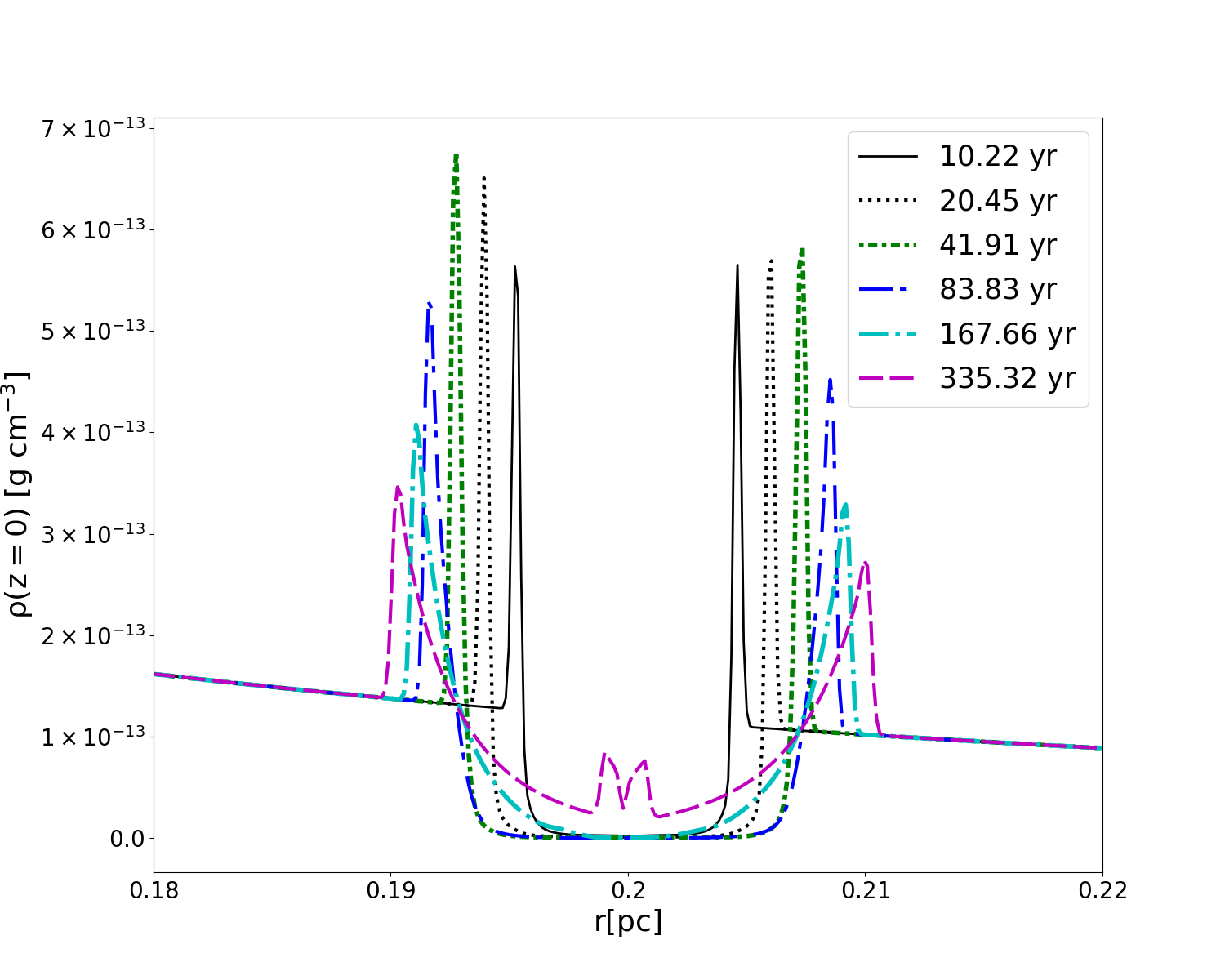}
\caption{Midplane density along a radial cut under $\phi=\pi/4$, in model 2 
($\gamma=1.1$).} 
\label{fig:density_model2}
\end{figure}

Now consider the mass that is contained below a height of $3H$ from the
midplane, i.e. at $z<3H$ (see also Figure \ref{fig:lost_mass_model1}). 
We see that the SN explosion is cleaning up material from $z<3H$ during the first 
$\sim 250$ yr after the explosion. It is interesting to note that 
a significant fraction of the mass that has been evacuated from
regions $z<3H$ remains in the region $3H<z<6.2H$. 
For instance, at $t=250$ yr, the mass evacuated from $z<3H$ is
$\sim 1000 M_{\odot}$ and $70\%$ of this mass remains at altitudes between 
$3H$ and $6.2H$.
Figure \ref{fig:density_averaged_phi} illustrates how the azimuthally-averaged
volume density, $\left<\rho\right>_{\phi}$, is distributed 
in the upper parts in the model 1 at two different times.

The velocity of expansion of the SNR depends on the adopted adiabatic index $\gamma$. 
The smaller the value of $\gamma$, the lower the internal pressure that pushes outward on
the shell. Because cooling in the accretion disk is
efficient, values of $\gamma$ close to $1$ are thought to be more realistic.
In order to quantify the dependence of
the flow pattern on the adiabatic index, we carried out a simulation with the
same parameters as model 1, but using $\gamma=1.1$ (model 2). 

\begin{figure}
\includegraphics[width=9.2cm,height=6.9cm]{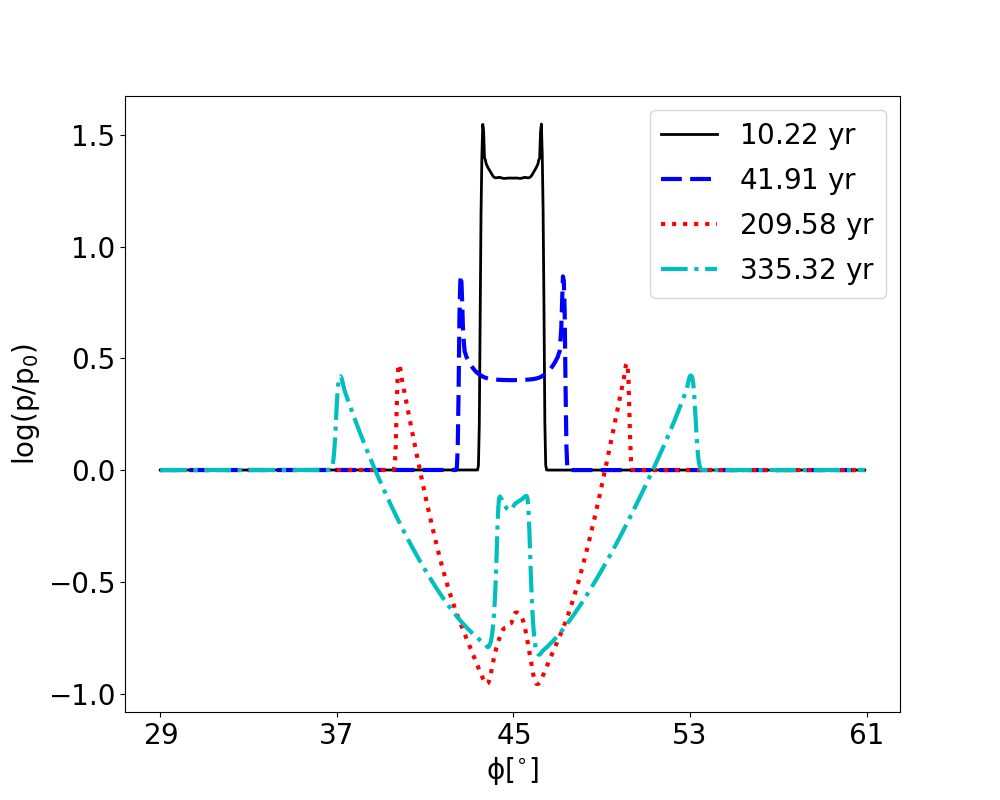}
\caption{Azimuthal distribution of the gas pressure at $z=0$ and $r=\RSNe$ in model 2 ($\gamma=1.1$). } 
\label{fig:pressure_model2}
\end{figure}

\begin{figure}
\includegraphics[height=7.2cm,width=9.2cm]{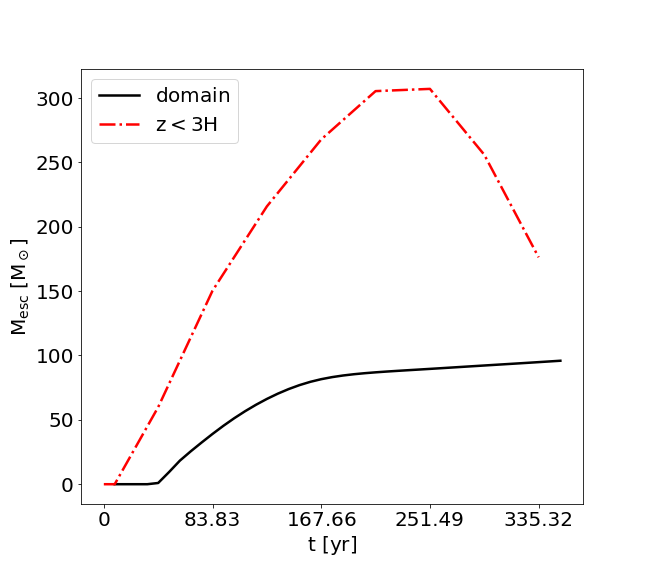}
\caption{Same as Figure \ref{fig:lost_mass_model1} but for model 2.} 
\label{fig:lost_mass_model2}
\end{figure}

Figure \ref{fig:density_model2} shows cuts of the density through the center 
of the SN explosion
along the radial direction in model 2, whereas Figure \ref{fig:pressure_model2} shows
cuts of the pressure along the azimuthal direction. The SNR acquires
a maximum width along the radial direction of $W_{\rm max}\simeq 0.02$ pc, which
is a factor of $1.3$ smaller than in model 1.  The expansion velocity of the SNR in the 
azimuthal direction is also slower than in model 1. The amount of mass that is carried outside the region $z<3H$ by the SNR is a factor of $\sim 3$ smaller in model 2 
than it is in model 1 (see Figure \ref{fig:lost_mass_model2}). Likewise, the mass 
that escapes from the computational domain is a factor of $3$ smaller in model 2 
than in model 1. In the next subsection, we will compare the amount of angular 
momentum that is redistributed by the SNR in models 1 and 2.

\subsection{Determing $\chi$ and $\beta$}
\label{sec:chi_beta}
In Sections \ref{sec:width} and \ref{sec:ang_L_redistribution}, we have introduced 
some dimensionless factors
when deriving the scaling laws that obey $W_{\rm max}$ and $\mathcal{J}_{\smalls \rm SNe}$ from theoretical grounds. These factors, which are expected to be of the order
of unity, can be measured in our simulations.

We have computed $\chi_{\smalls A}$ in our simulations as follows. From Equation
(\ref{eq:Rsh_scenarioA}),
the momentum imparted to the disk $\pi \tilde{\Sigma} R_{\rm sh}^{2} \dot{R}_{\rm sh}$
is $0.4 \pi \chi_{\smalls A} \tilde{\Sigma} \tilde{H}^{2} (\ESNe/\tilde{\rho}_{0} \tilde{H}^{3})^{1/2} =
2 \chi_{\smalls A} \tilde{H} (\ESNe \tilde{\Sigma})^{1/2}$. 
Therefore, we can measure $\chi_{\smalls A}$ in
our simulations as
\begin{equation}
\chi_{\smalls A} \simeq \frac{1} {\tilde{H} (\ESNe \tilde{\Sigma})^{1/2}} 
\int_{z<3H} \rho \,\delta v_{\smalls \parallel} d^{3}\vecr,
\end{equation}
where $\delta v_{\smalls \parallel}=\sqrt{\delta v_{x}^{2}+\delta v_{y}^{2}}$
is the planar component of the perturbed velocity field $\delta \vecv=\vecv-\vecv_{0}$,
with $\vecv_{0}$ is the unperturbed Keplerian velocity of the disk. In model 1,
we have computed $\chi_{\smalls A}$ at $t=41.9$ yr, when the low-density cavity 
produced by the explosion is still circular in the $z=0$ plane (see Figure \ref{fig:plane_dens_1}), and at $t=83.8$ yr, when the SNR has evacuated a 
significant mass of the disk. We find $\chi_{\smalls A} \simeq 0.7$ at both times.

Once we know $\chi_{\smalls A}$, we can evaluate the predicted maximum width of the 
SNR in scenario A as described in Section \ref{sec:width}. 
From Eqs. (\ref{eq:Wmax_A_trans_inner}) and (\ref{eq:Wmax_A_shear_inner}) with
$\chi_{\smalls A}=0.7$, $\xi=0.3$, $E_{51}=2$, we find 
$W_{\rm max}^{A,\rm trans}=0.02$ pc and $W_{\rm max}^{A,\rm shear}=0.011$ pc. 
The maximum width measured in model 1 is $0.026$ pc. 
Therefore, our $W_{\rm max}^{A,\rm shear}$ underestimates the width
by a factor of $\sim 2$. Interestingly, for $\chi_{\smalls A}=1$,
$W_{\rm max}^{A,\rm trans}=0.024$ pc, which is close to the value obtained from our simulations.

A value of $\chi_{\smalls A}\simeq 0.7$ is probably more adequate for simulations 
with a lower value of $\gamma$. 
In order to test this idea, we carried out a simulation
with all the same parameters except $E_{51}=78$ and $\gamma=1.4$.
We measured $W_{\rm max}=0.04$ pc in our simulation, which agrees
with $W_{\rm max}^{A,\rm trans}$. On the other hand,
$W_{\rm max}^{A,\rm shear}$ still underestimates the width by a factor of $2$.

\begin{figure}
\includegraphics[width=9.2cm,height=6.9cm]{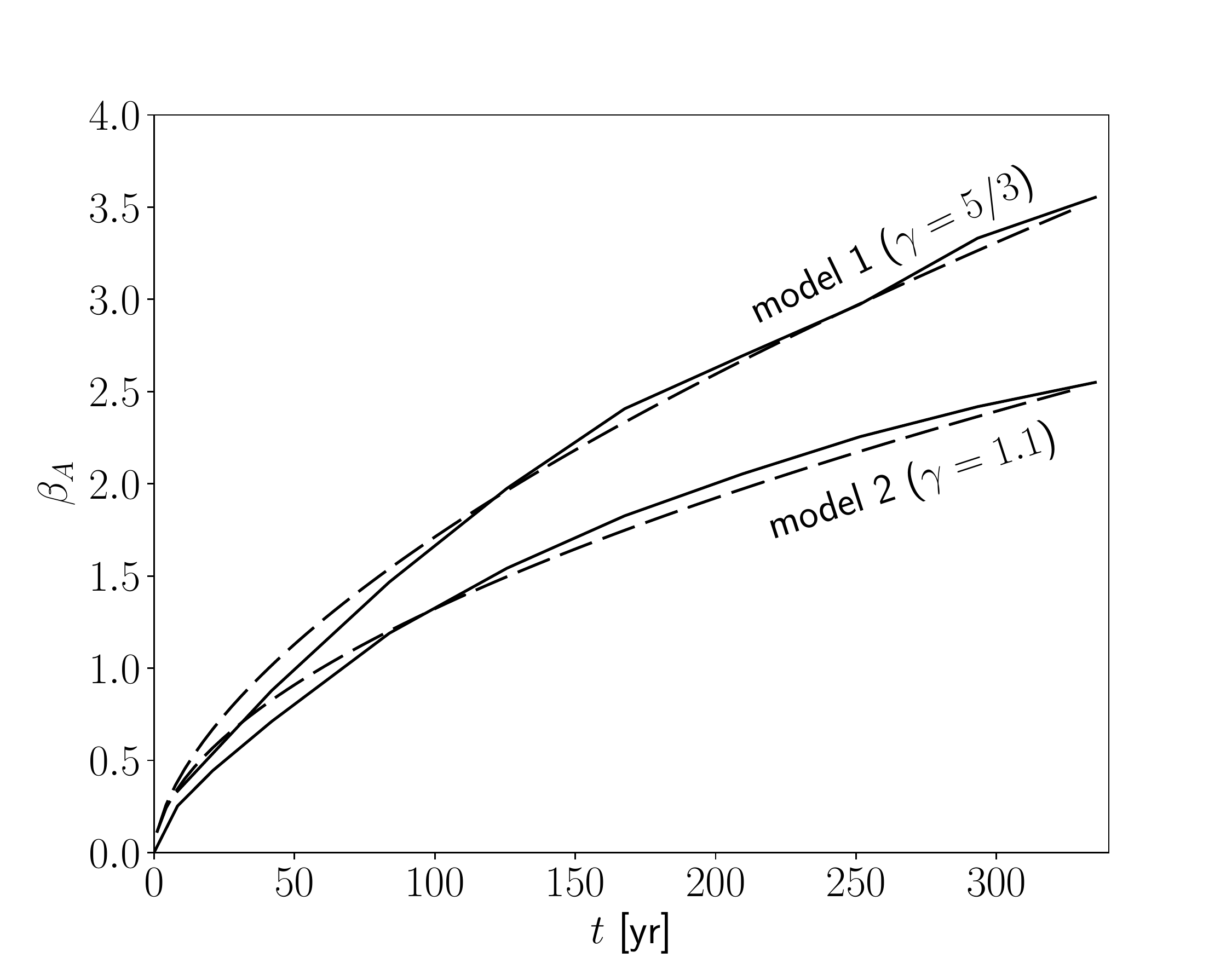}
\caption{$\beta_{\smalls A}$ vs. time (solid lines), overplotted with
the fits $\beta \propto t^{l}$ (dashed lines). The power-law indices are $l=0.6$ and
$l=0.54$ for model 1 and model 2, respectively.
$\beta_{A}$ is given in Equation (\ref{eq:betaA_sims}) and is a measurement of 
the transport of angular momentum by a single SN explosion.} 
\label{fig:beta_model1}
\end{figure}

In the following, we evaluate $\mathcal{J}_{\smalls \rm SNe}$, that is the amount
of angular momentum transported outward across a circle of radius $\RSNe$. 
More specifically,
we give $\beta_{\smalls A}$ (see \S \ref{sec:ang_L_redistribution}), which is related
to $\mathcal{J}_{\smalls \rm SNe}$ by
\begin{equation}
\beta_{\smalls A}=\frac{\mathcal{J}_{\rm \smalls SNe}}{0.6 (\ESNe \tilde{\Sigma})^{1/2}
\tilde{H}\RSNe}
\label{eq:betaA_sims}
\end{equation}
(see Equation \ref{eq:JJ_single_A}).
In order to determine $\beta_{\smalls A}$, we have measured 
${\mathcal{J}}_{\smalls \rm SNe}(t)$
in our simulations. The remainder of the variables in Equation 
(\ref{eq:betaA_sims}) are known input parameters. Since we are simulating the upper
half of the disk, we include a factor of $2$ in the calculation of 
${\mathcal{J}}_{\smalls \rm SNe}$. The resultant $\beta_{\smalls A}$ in models 1 and 2
are shown in Figure \ref{fig:beta_model1}. We see that $\beta_{\smalls A}$
increases over time following a power law.
In an inviscid disk, the value of $\beta_{\smalls A}$ is expected to converge asymptotically
to a constant value at large time. 2D simulations indicate that for the parameters
used in these simulations, the angular momentum ceases to drift outward
$\sim 1 P_{\rm orb}$ after the explosion. Consistent with this result,
we find that, in our 3D simulations, the rms azimuthal
component of perturbed velocity $\delta v_{\phi}$ in the shell decays with a characteristic 
timescale of $0.7P_{\rm orb}$. Using the power-law fits shown
in Figure \ref{fig:beta_model1}, we evaluate the value
of $\beta_{\smalls A}$ at $1P_{\rm orb}=838$ yr to obtain
that $\beta_{\smalls A}\simeq 6$
in model 1, and $\beta_{\smalls A}\simeq 4$ in model 2. We will use these estimates
of $\beta_{\smalls A}$ in Section \ref{sec:repeated_SN} when we study the effect
of repeated SN explosions in the accretion disk.

\begin{figure}
\includegraphics[width=0.5\textwidth,height=13 cm]{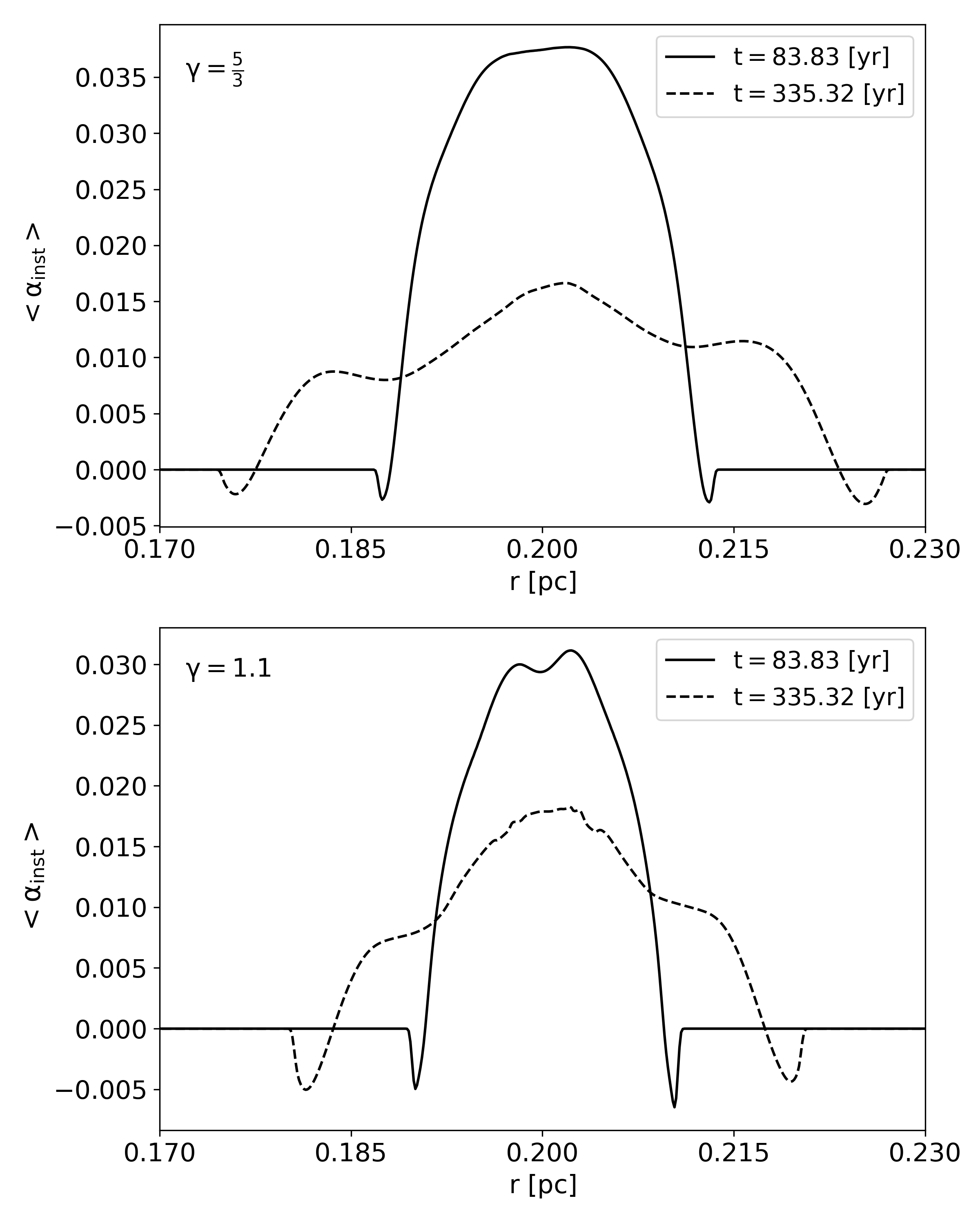}
\caption{Radial profile of $\left<\alpha_{\rm inst}\right>$ at 
$t=83.83$ yr and $t=335.3$ yr in model 1 (top) and in model 2 (bottom). The azimuthal average was done in our
computational domain $0\leq \phi \leq \pi/2$. Averaging between $0$ and
$2\pi$ would result in a lower $\left<\alpha_{\rm inst}\right>$ by a factor of $4$. } 
\label{fig:instant_alpha1}
\end{figure}

To illustrate how the angular momentum transport occurs in the midplane of the disk, 
we calculate 
\begin{equation}
\left<\alpha_{\rm inst}\right>\equiv \left<T_{R,\phi}\right>/\left<p\right>,
\end{equation}
where $T_{R,\phi}=\rho v_{R} \delta v_{\phi}$ 
is the Reynolds stress, $p$ is the gas pressure and
$\left<...\right>$ indicates averaging over $\phi$.
Figure \ref{fig:instant_alpha1} shows the $\left<\alpha_{\rm inst}\right>$ 
coefficient at the midplane for models 1 and 2 at two different times. 
Integrated over $R$, $\left<\alpha_{\rm inst}\right>$ is positive, signifying that the
net transport of angular momentum is outwards. 
We also observe that $\left<\alpha_{\rm inst}\right>$
at $R=\RSNe=0.2$ pc decreases over time.

\section{Repeated SN explosions: the viscosity parameter $\alpha$
in steady state}
\label{sec:repeated_SN}
In the previous Sections, we have calculated the 
amount of angular momentum that a single SN explosion can transport outwards. 
The effect of many SN explosions can be represented as an effective viscosity 
$\alpha_{\smalls \rm SNe}$. As said in \S \ref{sec:basis},  other agents 
besides SN explosions, may be at work in AGN accretion disks that contribute 
to the effective viscosity of the disk $\alpha$. The redistribution of angular
momentum by SN explosions will be the major contributor to the viscosity 
if $\alpha_{\smalls \rm SNe}\simeq \alpha$. Given the rate of SN explosions per
unit area in the disk $\dot{n}_{\smalls \rm SNe}$, we can evaluate
$\alpha_{\smalls \rm SNe}$. This will be done in the next Sections.

\subsection{Contribution of SN explosions to $\alpha$ }

We can always express the redistribution of angular momentum
through a radial flux $F_{J} (R,t)$. We will use the convention that $F_{J}(R,t)>0$ implies
that the transport of angular momentum is outwards. If this flux were carried by
density waves in a non-dissipative medium, then the angular momentum can be
transported to infinity without deposition into the disk. However, in our case,
the redistribution of angular momentum occurs at scales of the order of $W$.
Following \citet{goo01}, we may write the flux driven by SN explosions as
\begin{equation}
F_{J}(R) = 2\pi \int R_{\rm \smalls SNe} \dot{n}_{\rm \smalls SNe} \mathcal{J}_{\smalls {\rm SNe}} \,\phi(R-\RSNe)
\,d\RSNe,
\end{equation}
where $\phi(x)$ is a dimensionless distribution function with thickness 
$\sim 2W_{\rm max}$ and $\phi(0)=1$; it accounts for the damping length for
the deposition of angular momentum.
We may approximate the flux as
\begin{eqnarray}
F_{J}(R) && \simeq 4 \pi \dot{n}_{\rm \smalls SNe} W_{\rm max}R\,
\mathcal{J}_{\rm \smalls SNe}  
\label{eq:FJ_multiple_SN}
\end{eqnarray}
The expected rate of SN explosions can be estimated as
$\dot{n}_{\rm \smalls SNe}=f_{\rm \smalls SNe}\dot{\Sigma}_{\star}$, where $\dot{\Sigma}_{\star}$ is the star formation rate per unit area and
$f_{\rm \smalls SNe}$ is the number of massive stars per
solar mass of the star formation. To estimate $f_{\rm \smalls SNe}$, we 
assume a Salpeter initial mass function with the low mass cut-off at 
$0.1M_{\odot}$. We have $f_{\rm \smalls SNe}=0.01M_{\odot}^{-1}$.

In order to estimate the viscosity parameter, we consider that the angular 
momentum flux in a steady-state Keplerian disk has the form
\begin{equation}
F_{J}(R)= 3\pi \alpha_{\smalls \rm SNe} c_{s} H \Sigma R^{2} \Omega.
\label{eq:FJ_Shakura}
\end{equation}
By equating Eqs. (\ref{eq:FJ_multiple_SN}) and (\ref{eq:FJ_Shakura}), we find
\begin{equation}
\alpha_{\smalls \rm SNe}\simeq \frac{4}{3} f_{\rm \smalls SNe} \left(\frac{W_{\rm max}}{R}\right)
\left(\frac{\dot{\Sigma}_{\star}}{\Sigma}\right) 
\left(\frac{\mathcal{J}_{\rm \smalls SNe}}{c_{s}^{2}}\right).
\label{eq:alphaSNe_general}
\end{equation}
In the next Section, we estimate $\alpha_{\smalls \rm SNe}$ in the SG model.
To do so, we need the star formation rate $\dot{\Sigma}_{\star}$.

\subsection{Application to the SG model}
\label{sec:SNe_area}
 
Assuming that stellar feedback is the main agent 
to provide vertical support to 
the disk, we can infer the required star formation rate per unit area, denoted by
$\dot{\Sigma}_{\star}^{\smalls \rm sup}$ (see Appendix \ref{sec:TQM_appendix}).
Figure \ref{fig:SFR_profile} shows $\dot{\Sigma}_{\star}^{\smalls \rm sup}$
for our fiducial parameters. As discussed in TQM, the star formation rate has a bump
where the opacity is low (in the `opacity gap'). In our case, the disk becomes optically
thin at $R_{3}>10$.

As anticipated in \S \ref{sec:basis} and discussed in detail in TQM, a large
$\dot{\Sigma}_{\star}^{\smalls \rm sup}$ makes it difficult to fuel the central SMBH,
as star formation may halt gas accretion onto the central SMBH (starvation).
More specifically, star formation produces starvation if 
\begin{equation}
\dot{M}_{\rm acc}\leq 2\pi (1+f_{\smalls \rm SNe} M_{\small \rm exp})
\int_{R_{\rm min}}^{R_{\rm max}} \dot{\Sigma}_{\star} R\; dR,
\label{eq:starvation}
\end{equation}
where $R_{\rm min}=10^{3}R_{\smalls \rm Sch}$, 
$R_{\rm max}=10^{5}R_{\smalls \rm Sch}$, $M_{\small \rm exp}$ is the mass 
expelled from the disk to the corona by one SN explosion.

Consider the phenomenological Kennicutt-Schmidt law 
$\dot{\Sigma}_{\star}=C_{\small \rm KS} \Sigma^{7/5}$,
where $\Sigma$ is in unit of $M_{\odot}$pc$^{-2}$ and $\dot{\Sigma}_{\star}$ in
$M_{\odot}$pc$^{-2}$yr$^{-1}$ \citep{ken98,che09}. Using Equations (\ref{eq:dotMacc}),
(\ref{eq:initial_Sigma}) and (\ref{eq:starvation}), it is easy to show that in order to avoid
starvation we need $C_{\rm KS}< C_{\rm cr}$, where 
\begin{equation}
C_{\small \rm cr}= 1.5 \times 10^{-10} \alpha_{\smalls 0.1} 
g_{\smalls \rm SNe}^{-1} Q_{\rm m}^{0.93}\xi^{0.53} M_{8}^{-0.07},
\end{equation}
with $g_{\smalls \rm SNe}\equiv 1+f_{\smalls \rm SNe}M_{\smalls \rm exp}$.

In Figure \ref{fig:SFR_profile} we can compare $\dot{\Sigma}_{\star}^{\rm cr}\equiv
C_{\rm cr}\Sigma^{7/5}$  
and $\dot{\Sigma}_{\star}^{\rm sup}$. We clearly
see that $\dot{\Sigma}_{\star}^{\rm cr}$ is not enough to
provide support to the disk and additional sources of feedback/heating are required.

\begin{figure}
\includegraphics[scale=0.38]{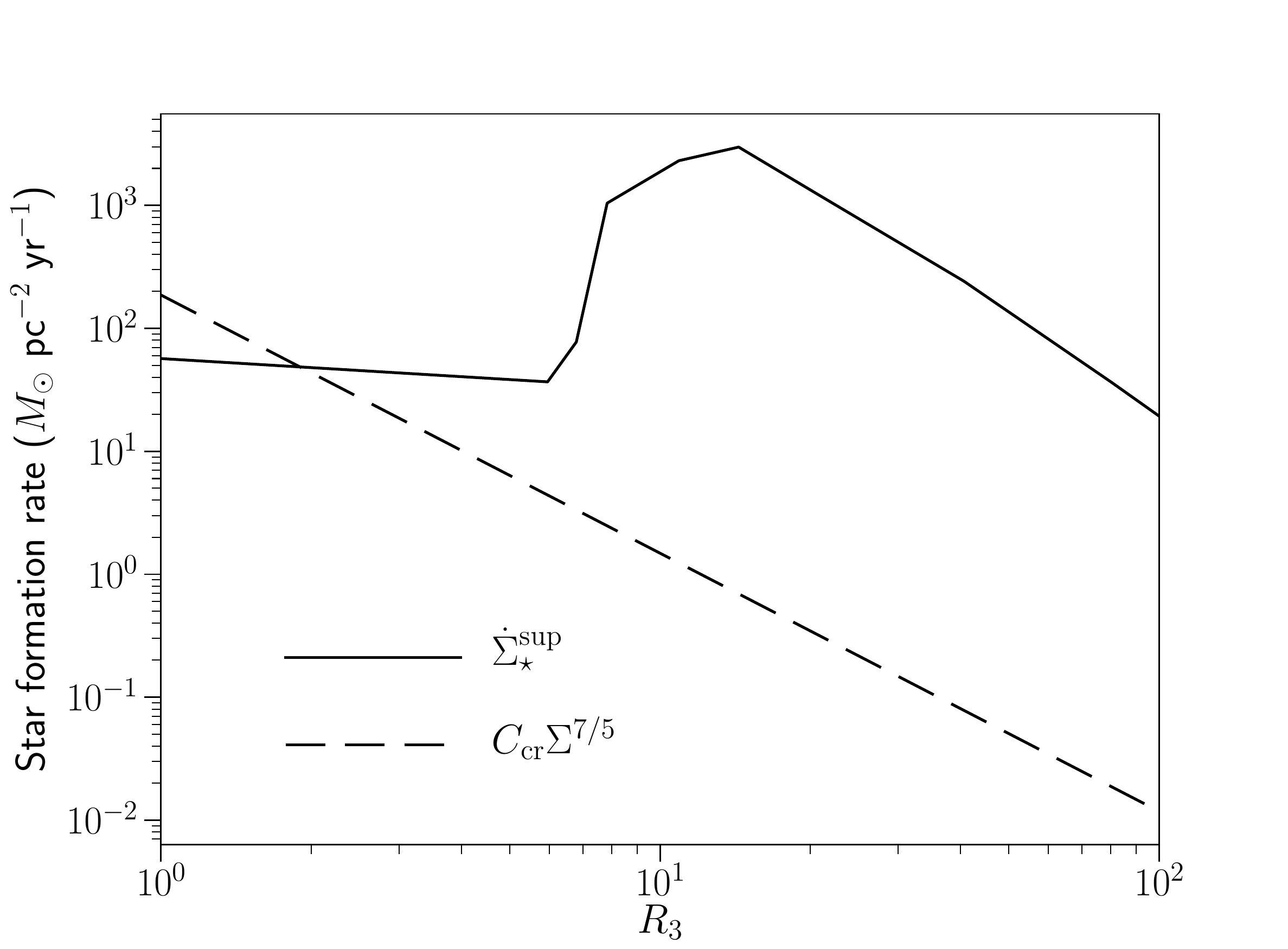}
\caption{Radial profile of $\dot{\Sigma}_{\star}^{\smalls \rm sup}$ (solid line)
and a Kennicutt-Schmidt law $C_{\rm cr}\Sigma^{7/5}$ (dashed line). We have used
the following parameters:
$M_{8}=1$, $\alpha_{\smalls 0.1}=1$, $\xi=0.3$, $Q_{\rm m}=1$ and 
$M_{\smalls \rm exp}=100M_{\odot}$.} 
\label{fig:SFR_profile}
\vskip 0.3cm
\end{figure}

Substituting Eq. (\ref{eq:JJ_single_A}) into Eq. (\ref{eq:alphaSNe_general}),
and generously assuming 
$\dot{\Sigma}_{\star}=\dot{\Sigma}_{\star}^{\rm cr}$,
we get the following upper value
\begin{equation}
\alpha^{\smalls (A)}_{\smalls \rm SNe}\simeq 1\times 10^{-3}\beta_{\smalls A} 
\xi^{0.53}\alpha_{\smalls 0.1}g_{\smalls \rm SNe}^{-1} M_{8} E_{51}^{0.5}  R_{3}^{1.15} \left(\frac{W_{\rm max}}{R}\right)
\label{eq:alphaA}
\end{equation}
in scenario A. In the above equation we have used 
$f_{\rm \smalls SNe}=0.01 M_{\odot}^{-1}$.

On the other hand, combining Eq. (\ref{eq:Delta_J_roz}) and Eq.  
(\ref{eq:alphaSNe_general}), we arrive at the following equation
\begin{equation}
\alpha^{\smalls (B)}_{\smalls \rm SNe}\simeq 1.3\times 10^{-3} \beta_{\smalls B}
\alpha_{\smalls 0.1}g_{\smalls \rm SNe}^{-1}
\wh{M}_{10}^{1/2}E_{51}^{1/2} R_{3}^{2/5} \left(\frac{W_{\rm max}}{R}\right)
\label{eq:alphaB}
\end{equation}
in scenario B. We recall that the equations above are valid as long as 
$\alpha_{\smalls \rm SNe}\leq \alpha$.

We see that the upper values for $\alpha^{\smalls (A)}_{\smalls \rm SNe}$ and
$\alpha^{\smalls (B)}_{\smalls \rm SNe}$ do not explicitly depend on the precise
value adopted for $Q_{\rm m}$, but they depend weakly on $Q_{\rm m}$ through
$W_{\rm max}$ (see, e.g., Eqs. \ref{eq:Wmax_A_trans_model1},
\ref{eq:Wmax_A_shear_inner_model1}, \ref{eq:Wmax_trans_B} and
\ref{eq:Wmax_shear_B}).

Now, we evaluate $\alpha^{\smalls (A)}_{\smalls \rm SNe}$ for our reference values
$\alpha_{\smalls 0.1}=1$, $Q_{\rm m}=1$, $M_{8}=1$, $\xi=0.3$, $E_{51}=2$ 
and $\gamma=5/3$.
For these parameters, we have found in Section \ref{sec:simulations} that 
$\beta_{\smalls A}\simeq 6$, $M_{\smalls \rm \exp}=300M_{\odot}$ and
$W_{\rm max}\sim 0.1$ pc.
For these values, Equation (\ref{eq:alphaA}) implies $\alpha^{\smalls (A)}_{\smalls \rm SNe}\simeq 0.02$ and, therefore, SN explosions may contribute up to $\sim 20\%$ to
the effective viscosity in this model.

For $\gamma=1.1$, $\alpha_{\smalls 0.1}=1$, $M_{8}=1$, $\xi=0.3$, $E_{51}=2$,
$\beta_{\smalls A}\simeq 4$, $M_{\smalls \rm exp}=100M_{\odot}$ and
$W_{\rm max}\sim 0.075$ pc, we find a similar value of 
$\alpha^{\smalls (A)}_{\smalls \rm SNe}\simeq 0.02$. 

Interestingly, $\alpha^{\smalls (A)}_{\smalls \rm SNe}$ increases with $M_{8}$.
More specifically, since $W_{\rm max}^{\smalls A,\rm shear} \propto M_{8}^{8/9}$ 
(Equation \ref{eq:Wmax_A_shear_inner_model1}) and $R\propto M_{8}R_{3}$, 
Equation (\ref{eq:alphaA}) predicts 
$\alpha_{\smalls \rm SNe}^{\smalls (A)}\propto M_{8}^{8/9}$. Therefore, 
$\alpha_{\smalls \rm SNe}^{\smalls (A)}\simeq \alpha$ if $M_{8}\simeq 5$.

In scenario B, we do not have an estimate of $\beta_{\smalls B}$ but certainly
$\beta_{\smalls B}\leq 1$ (see Appendix \ref{sec:scenarioB}). If we generously take 
$\beta_{\smalls B}=g_{\smalls \rm SNe}=1$,
Equation (\ref{eq:alphaB}) with $\alpha_{\smalls 0.1}=1$, 
$M_{8}=1$, $E_{51}=2$, $\wh{M}_{10}=1$, $W_{\rm max}/R\sim 0.05$, 
$R_{3}^{2/5}\simeq 6.5$ and $\xi=0.3$, implies
$\alpha^{\smalls (B)}_{\smalls \rm SNe}\simeq 6\times 10^{-4}$. This small
value of $\alpha^{\smalls (B)}_{\smalls \rm SNe}$ compared to $\alpha$ means
that SN explosions are not efficient to drive the effective viscosity in this model
and other sources need to be invoked.

In scenario B, the effective $\alpha_{\smalls \rm SNe}$ viscosity depends weakly on $M_{8}$. 
For illustration, 
consider a disk around a black hole with $M_{8}\ll 1$. In this case we are in the limit
$\Delta \gg H$ (Appendix \ref{sec:limit1}). From Equation (\ref{eq:Wmax_shear_limit1}) 
with $\chi'\propto M_{8}$, we have 
$W_{\rm max}\propto M_{8}^{8/9}$. Since $R\propto M_{8}R_{3}$, we get
$\alpha_{\smalls \rm SNe}^{\smalls (A)}\propto (W_{\rm max}/R)\propto M_{8}^{-1/9}$.

Our result that $\alpha_{\smalls \rm SNe}^{\smalls (B)}\simeq 6\times 10^{-4}$ 
for $\xi=0.3$, $M_{8}=1$ and a SN rate of $8\times 10^{-3}$ yr$^{-1}$, contrasts 
with the findings of \citet{roz95} that $\alpha_{\smalls \rm SNe}\simeq 0.1$ 
for $M_{8}=1$ and a SN rate of $10^{-4}$ yr$^{-1}$ in scenario B.
\citet{roz95} overestimated $\alpha_{\smalls \rm SNe}$ because they implicitly assumed that
the flow carries the angular momentum without any dissipation. A finite
damping length leads to a smaller effective viscosity, since angular momentum
is deposited in the disk.

\section{Summary and conclusions}
\label{sec:conclusions}
We have studied the role of SN explosions on the density structure and angular 
momentum redistribution in AGN accretion disks within a $1$ pc scale.
In our models, the AGN accretion disk properties are taken from the accretion disk
model derived by SG. A SN explosion drives a shock that sweeps up mass,
forms a cavity in the disk and redistributes disk angular momentum.
We have provided some analytical estimates
of the width of the cavity and the redistribution of angular momentum induced by
a single SN explosion. We have introduced 
some fudge factors to include deviations from our simple approximations.
By means of 3D hydrodynamical simulations, which take into account
the lost mass and the momentum carried by the outflow,
we have calibrated these fudge factors.

The radial width of the cavity, the mass ejected from the disk and the amount of 
angular momentum that is redistributed in the disk ($\mathcal{J}_{\smalls \rm SNe}$)
by a single SN explosion all depend upon the value adopted for the adiabatic index 
$\gamma$ of the gas. As a reference number, for $\gamma=1.1$, we find $\mathcal{J}_{\smalls \rm SNe}\sim 8\times 10^{60}$ g cm$^{2}$ s$^{-1}$
for a SN explosion at a radius $0.2$ pc, in our disk model with $M_{8}=1$. 

We have estimated the effective $\alpha_{\smalls \rm SNe}$ viscosity provided by SN explosions in a steady state, where
the rate of SN explosions is determined by adopting a Kennicutt-Schmidt law for the
star formation. For $\gamma$ between $1.1$ and $5/3$, we find that 
$\alpha_{\smalls \rm SNe}\gtrsim 0.1$ if $M_{8}\gtrsim 5$.

Some authors adopted the momentum conservation limit to infer 
$\alpha_{\smalls \rm SNe}$ \citep[e.g.,][]{roz95,col99}.
In this limit, which is relevant when 
cooling is already important in the final stages of the free expansion
phase, we find $\alpha_{\smalls \rm SNe}/\alpha\lesssim 6\times 10^{-3}$. 
Therefore, the contribution of SN explosions to the effective viscosity is negligible.

We have assumed that SN explosions occur in a smooth disk. However, unless the stellar heating is effective,
the disk may fragment into clouds through gravitational instabilities \citep[e.g.,][]{jia11}. 
If SN explosions occur inside dense clouds, part of their initial momentum will be 
absorbed by their natal clouds. Therefore, our estimates of $\alpha_{\smalls \rm SNe}$ 
should be treated as upper limits.

\acknowledgments
We thank the anonymous reviewer for helpful and constructive comments.
AM-B  and PFV acknowledge financial support from DGAPA-PAPIIT (UNAM) grant IG100218. FJS-S thanks PAPIIT for financial support through the project IN111118. 
ROC acknowledges postdoctoral CONACyT grant.

\appendix
\section{A. Pressure distribution and star formation rate}
\label{sec:TQM_appendix}
Following TQM, we can make predictions about the relative importance of
thermal, radiation and turbulent pressure and their dependence with distance.
We can also calculate the star formation rate per unit area. The equation of
vertical equilibrium (Equation 39 in TQM) becomes
\begin{equation}
\rho_{0}c_{s}^{2} = p_{\rm th} + \mathcal{E} \dot{\Sigma}_{\star}^{\smalls \rm sup} c \left(\frac{\tau}{2} +\lambda\right),
\end{equation}
where $\mathcal{E}$ is the efficiency with which star formation converts rest mass
into radiation and $\lambda$ is a dimensionless parameter that measures
the amount of the momentum injected by stars that is converted into turbulent
motions ($c$ is the light speed). To compute $p_{\rm th}$, we evaluate
the gas temperature $T$ through
\begin{equation}
T^{4}=\frac{3}{4} T_{\rm eff}^{4} \left(\tau + \frac{2}{3\tau} +\frac{4}{3}\right),
\end{equation}
where $\tau=\kappa \Sigma/2$, $\kappa$ is the disk opacity and
\begin{equation}
\sigma_{\rm SB} T_{\rm eff}^{4}= \frac{1}{2}\mathcal{E} \dot{\Sigma}_{\star}^{\smalls \rm sup}
c^{2}+\frac{3}{8\pi} \dot{M}_{\rm acc} \Omega^{2}.
\end{equation}
Here $\sigma_{\rm SB}$ is the Stefan-Boltzmann constant. 
Assuming ${\mathcal{E}}=10^{-3}$, $\lambda=1$, and the disk opacities 
given in TQM, which are based on \citet{sem03}, we can derive
$\dot{\Sigma}_{\star}^{\smalls \rm sup}$, $p_{\rm th}$, 
$p_{\rm rad}$ and $p_{\rm tur}$, in our disk model. 
We note that 
$p_{\rm rad}=\mathcal{E} \tau \dot{\Sigma}_{\star}^{\smalls \rm sup} c/2$ and 
$p_{\rm tur}=\lambda \mathcal{E} \dot{\Sigma}_{\star}^{\smalls \rm sup} c$.
To easy comparison, we have taken the same values for $\mathcal{E}$ and
$\lambda$ as TQM, but we warn that they are uncertain.

\section{B. Maximum radial width of the SNR in the limit $\Delta \gg \tilde{H}$}
\label{sec:limit1}
As pointed out in Section \ref{sec:free_expansion}, in some cases (especially for $M_{8}<1$),  
the SN explosion can break out of the disk in the free expansion phase. 
In cases in which $\Delta \gg \tilde{H}$, this free expansion phase
continues until $R_{\rm sh}\simeq \Delta$. We assume that a momentum 
conservation phase begins just after the free expansion phase ends. 
Only the momentum of those elements ejected in a polar angle $\hat{\theta}$ between
$\hat{\theta}_{\rm min}\simeq \arccos (\tilde{H}/(\sqrt{\Delta^{2}+\tilde{H}^{2}})$ and
$\hat{\theta}_{\rm max}\simeq \pi - \arccos (\tilde{H}/(\sqrt{\Delta^{2}+\tilde{H}^{2}})$
will be able to push the disk along the planar directions. 
Thus, the momentum absorbed by the disk will be
\begin{equation}
P_{\rm abs}\simeq \frac{P_{\smalls \rm SNe}}{4\pi} \int_{0}^{2\pi}
\int_{\hat{\theta}_{\rm min}}^{\hat{\theta}_{\rm max}} \sin^{2}\hat{\theta} \,d\theta
\, d\phi=P_{\smalls \rm SNe} \frac{\tilde{H}}{\sqrt{\Delta^{2}+\tilde{H}^{2}}}.
\end{equation}
It is easy to show that the mass that escapes from the disk is 
$\simeq 0.5\pi \tilde{\Sigma}\Delta^{2}$. 
Momentum conservation implies that $\dot{R}_{\rm sh}$ obeys
\begin{equation}
\pi \tilde{\Sigma}\left(R_{\rm sh}^{2}-\frac{\Delta^{2}}{2}\right) \dot{R}_{\rm sh}=
\chi' P_{\smalls\rm SNe},
\end{equation}
with $\chi'\equiv \tilde{H}/\sqrt{\Delta^{2}+\tilde{H}^{2}}$.
The transonic condition (see \S \ref{sec:width}) occurs when the width of the SNR in the radial direction is 
\begin{equation}
W_{\rm max}^{\smalls \rm trans}\simeq 2\left(
\frac{\chi' P_{\smalls \rm SNe}}{\pi \tilde{\Sigma} 
\tilde{c}_{s}}+
\frac{\Delta^{2}}{2}\right)^{1/2}.
\end{equation}

On the other hand, $W_{\rm max}^{\smalls \rm shear}$ satisfies a cubic equation.
Here we just provide a lower limit:
\begin{equation}
W_{\rm max}^{\smalls \rm shear}> 1.5
\left(\frac{\chi' P_{\smalls \rm SNe}}{\tilde{\Sigma}\tilde{\Omega}}\right)^{1/3}.
\end{equation}
We notice that $W_{\rm max}^{\smalls \rm trans}$ and 
$W_{\rm max}^{\smalls \rm shear}$ do not depend explicitly
on $\ESNe$ because the adiabatic phase never develops. 

In our disk model (Eqs. \ref{eq:initial_Sigma}-\ref{eq:vrms}),  
approximating $\chi'\simeq \tilde{H}/\Delta$, the widths read
\begin{equation}
W_{\rm max}^{\smalls \rm trans} = 10^{-3} \xi^{-1/12} Q_{\rm m}^{1/3}M_{8}^{2/3}
\wh{M}_{10}^{1/2} E_{51}^{1/4} \tilde{R}_{3}^{9/8} (1+4\times 10^{-3} \xi^{-1/6}
Q_{\rm m}^{1/3}M_{8}^{-2/3} \tilde{R}_{3}^{-3/4})^{1/2}\,  {\rm pc},
\label{eq:Wmax_trans_limit1}
\end{equation}
and
\begin{equation}
W_{\rm max}^{\smalls \rm shear} > 4.5\times 10^{-4} \chi'^{1/3}\xi^{-1/9} 
Q_{\rm m}^{2/9} M_{8}^{5/9} \wh{M}_{10}^{1/6} E_{51}^{1/6}\tilde{R}_{3}\, {\rm pc}.
\label{eq:Wmax_shear_limit1}
\end{equation}

\section{C. Maximum radial width of the SNR in scenario B}
\label{sec:scenarioB}
Scenario B assumes that the internal pressure in the cavity is negligible and, 
therefore, the SNR evolves as a pure momentum-driven snowplow. We will distinguish between cases where the SN explosion occurs within the radius $\tilde{R}_{b}^{\smalls (B)}$ 
and beyond $\tilde{R}_{b}^{\smalls (B)}$.

{\bf Case $\tilde{R}_{3}>\tilde{R}_{b}^{\smalls (B)}$.-} Since the SNR cannot breakout of the disk,
it will hardly reach a height larger than $\simeq \sqrt{3}\tilde{H}$.
Momentum-conservation dictates
\begin{equation}
P_{\rm \smalls SNe}\simeq \frac{4\pi}{3}\tilde{\rho}_{0} R_{\rm sh}^{3} \dot{R}_{\rm sh}.
\end{equation}
The transonic and shear conditions imply
\begin{equation}
W_{\rm max}^{\smalls B, \rm trans}\simeq 1.2
\left(\frac{P_{\rm \smalls SNe}}{\tilde{\rho}_{0} \tilde{c}_{s}}\right)^{1/3},
\end{equation}
and
\begin{equation}
W_{\rm max}^{\smalls B, \rm shear}\simeq 1.5\left(\frac{P_{\rm \smalls SNe}}{\tilde{\rho}_{0} \tilde{\Omega}}\right)^{1/4}.
\end{equation}
In terms of $M_{8}$ and $\tilde{R}_{3}$,
\begin{equation}
W_{\rm max}^{\smalls B, \rm trans}=5\times 10^{-4} \xi^{-1/9}Q_{\rm m}^{2/9}
M_{8}^{5/9}
\wh{M}_{10}^{1/6} E_{51}^{1/6}\tilde{R}_{3} \,{\rm pc},
\end{equation}
and 
\begin{equation}
W_{\rm max}^{\smalls B, \rm shear}=2\times 10^{-4} Q_{\rm m}^{1/4}M_{8}^{3/4}
\wh{M}_{10}^{1/8}E_{51}^{1/8} \tilde{R}_{3}^{9/8}\, {\rm pc}.
\end{equation}

{\bf Case $\tilde{R}_{3}<\tilde{R}_{b}^{\smalls (B)}$.-}
If the explosion site lies at a radius less than $\tilde{R}_{b}^{\smalls (B)}$, the SNR is able to break out of the disk. After breakout, i.e. when $R_{\rm sh}\gtrsim \sqrt{3}\tilde{H}$, the
shock velocity is given by 
\begin{equation}
\chi_{\smalls B} P_{\rm \smalls SNe}=\pi R_{\rm sh}^{2} \tilde{\Sigma}
\dot{R}_{\rm sh}.
\label{eq:scenarioB_dotR_R2}
\end{equation}
Here $\chi_{\smalls B}$ is the fraction of momentum that is absorbed by the disk;
thus $\chi_{\smalls B}\leq 1$.
Imposing the transonic and shear conditions, it follows that
\begin{equation}
W_{\rm max}^{\smalls B,\rm trans}\simeq \left(\frac{\chi_{\smalls B} P_{\rm \smalls SNe}}
{\tilde{\Sigma} \tilde{c}_{s}}\right)^{1/2},
\end{equation}
and
\begin{equation}
W_{\rm max}^{\smalls B,\rm shear}\simeq 1.5
\left(\frac{\chi_{\smalls B} P_{\rm \smalls SNe}}{\tilde{\Sigma}
\tilde{\Omega}}\right)^{1/3}.
\end{equation}
For our accretion disk model (Eqs. \ref{eq:initial_Sigma}-\ref{eq:vrms}),
\begin{equation}
W_{\rm max}^{\smalls B,\rm trans}\simeq 10^{-3} \chi_{\smalls B}^{1/2}\xi^{-1/3} 
Q_{\rm m}^{1/6}M_{8}^{1/6} \wh{M}_{10}^{1/4} E_{51}^{1/4}\tilde{R}_{3}^{3/4}\, {\rm pc},
\label{eq:Wmax_trans_B}
\end{equation}
and
\begin{equation}
W_{\rm max}^{\smalls B,\rm shear}\simeq 4.5\times 10^{-4} 
\chi_{\smalls B}^{1/3}\xi^{-1/9} Q_{\rm m}^{2/9}
M_{8}^{5/9} \wh{M}_{10}^{1/6} E_{51}^{1/6}\tilde{R}_{3}\, {\rm pc}.
\label{eq:Wmax_shear_B}
\end{equation}

\end{document}